\documentclass[twocolumn]{IEEEtran}

\usepackage{bm}
\usepackage{enumitem}
\usepackage{commath}
\usepackage{graphicx}
\graphicspath{{Pictures/}}
\usepackage{amsmath}
\usepackage{subcaption}
\usepackage{breqn}
\usepackage{cite}
\usepackage[table]{xcolor}
\usepackage{booktabs}
\usepackage{empheq}
\usepackage{amsfonts}
\usepackage{amssymb}
\usepackage{amsthm}
\usepackage{hyperref}
\usepackage{xcolor}
\usepackage{mathrsfs}
\usepackage{accents}
\usepackage{thmtools}
\usepackage{thm-restate}
\usepackage{float}
\usepackage{comment}
\usepackage{algpseudocode}
\usepackage{algorithm}

\newtheorem{lem}{Lemma}
\newtheorem{definition}{Definition}
\newtheorem{assumption}{Assumption}
\newtheorem{cor}{Corollary}
\newtheorem{proposition}{Proposition}
\newtheorem{remark}{Remark}
\newtheorem{example}{Example}
\newtheorem{problem}{Problem}
\newtheorem{thm}{Theorem}

\newcommand{\redtext}[1]{{\color{red}#1}}
\newcommand{\ubar}[1]{\underaccent{\bar}{#1}}

\newcommand{\myvar}[1]{\bm{#1}}

\newcommand{\barvar}[1]{\bar{\bm{#1}}}

\newcommand{\myset}[1]{\mathcal{#1}}

\newcommand{\mysetbound}[1]{\partial \mathcal{#1}}

\newcommand{\robvar}[1]{\hat{\myvar{#1}}}

\newcommand{\robset}[1]{\hat{\myset{#1}}}

\newcommand{\Lip}[2]{\mathcal{L}_{#1}^{#2}}

\newcommand{\ASBF}{\emph{asymptotically stable discrete-time barrier function} }
\newcommand{\RBF}{\emph{robust, discrete-time barrier function}}
\newcommand{\ASRBF}{\emph{robust, asymptotically stable discrete-time barrier function} }

\begin{document}


\title{
A Robust, Efficient Predictive Safety Filter 
}

\author{Wenceslao Shaw Cortez, Jan Drgona, Draguna Vrabie, Mahantesh Halappanavar 
\thanks{\textsuperscript{\dag} W. Shaw Cortez, J. Drgona, D. Vrabie, and M. Halappanavar are with the Data Science and Machine Intelligence Department at the Pacific Northwest National Laboratory, Richland, WA, USA. 
        E-Mail: {\tt\small $\{$w.shawcortez, jan.drgona, draguna.vrabie, mahantesh.halappanavar $\}$@pnnl.gov}}
\thanks{This research was partially supported by the U.S. Department of Energy, through the Office of Advanced Scientific Computing Research's “Data-Driven Decision Control for Complex Systems (DnC2S)” project, and through the Energy Efficiency and Renewable Energy, Building Technologies Office under the “Advancing Market-Ready Building Energy Management by Cost-Effective Differentiable Predictive Control” projects. PNNL is a multi-program national laboratory operated for the U.S. Department of Energy (DOE) by Battelle Memorial Institute under Contract No. DE-AC05-76RL0-1830.}
}

\maketitle
\thispagestyle{empty}
\pagestyle{empty}

\begin{abstract}
In this paper, we propose a novel predictive safety filter that is robust to bounded perturbations and is implemented in an even-triggered fashion to reduce online computation. The proposed safety filter extends upon existing work to reject disturbances for discrete-time, time-varying nonlinear systems with time-varying constraints. The safety filter is based on novel concepts of robust, discrete-time barrier functions and can be used to filter any control law. Here, we use the safety filter in conjunction with Differentiable Predictive Control (DPC) as a promising offline learning-based policy optimization method. The approach is demonstrated on a two-tank system, building, and single-integrator example.
\end{abstract}


\section{Introduction}

Control barrier functions are a useful tool for ensuring constraint satisfaction of nonlinear, dynamical systems \cite{Ames2017}. This method has received increasing attention recently due to its modularity and applicability to learning-based techniques that generally do not have guarantees of constraint satisfaction, i.e., safety~\cite{Nghiem2023,SafeControlReview2022}. Barrier functions are also advantageous for being robust to perturbations and provide asymptotic stability to the safe set. 

One criticism of barrier function methods, however, is the difficulty in constructing the functions themselves. \cite{Freire2023} provides a way to construct barrier functions using maximal output admissible sets, but is dependent on a known stabilizing control law. Verification methods have been developed for uncertain system dynamics \cite{Akella2022a}, but only probabilistic guarantees of safety are provided. Some synthesis methods are restricted to specific types of systems \cite{ShawCortez2022a} or are dependent on sampling methods or sum-of-squares techniques, which do not scale well with system size \cite{Clark2021}. Other approaches require expert-provided trajectories in order to learn the barrier function \cite{robey2020,Kehan2021}. Most of these approaches require significant offline data and computation, may be subject to conservatism, and are generally restricted to time-invariant systems. An alternative approach considered here is to use a prediction horizon to relax these restrictions.

\begin{figure}[t!]
\centering
  \includegraphics[width=1.0\linewidth]{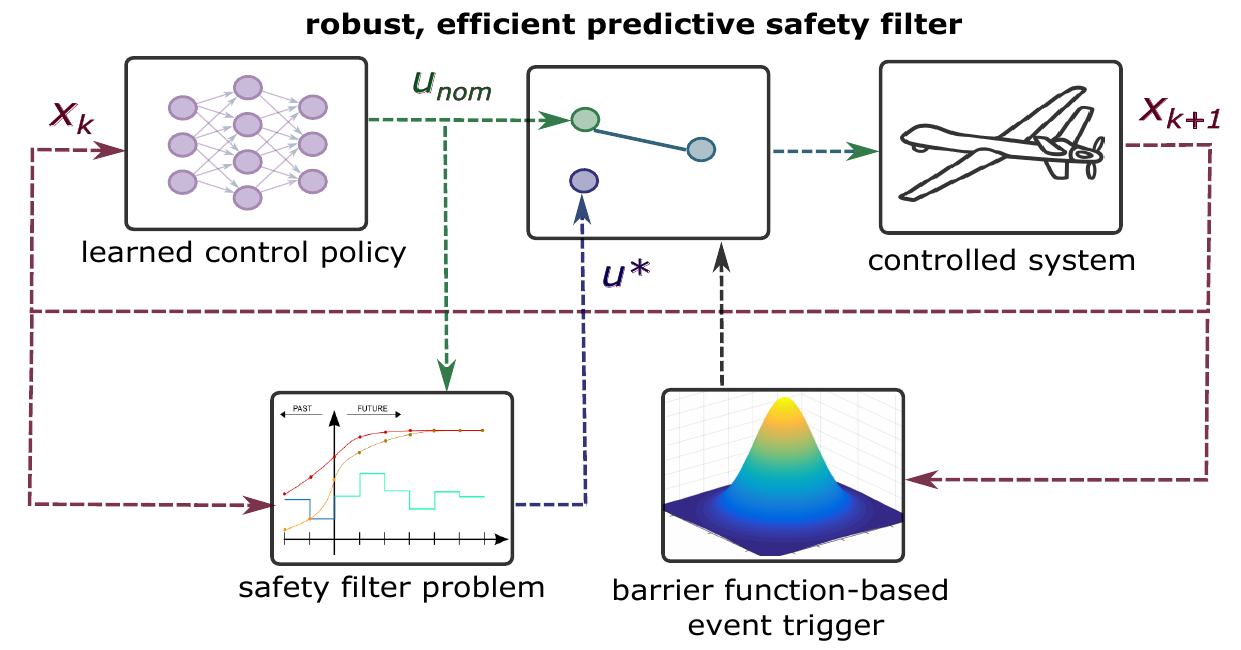}
  \vspace{-0.4cm}
    \caption{Schematics of the proposed event-triggered robust predictive safety filter combined with learning-based policy.}
    \label{fig:graphical_abstract}
\end{figure}

Many existing barrier function methods implement the safety-critical control as a nonlinear program, which is effectively a 1-step look-ahead model predictive control (MPC) problem. The ability to predict system behavior can relax conservatism in the 1-step look-ahead approach. There exist methods that combine the finite horizon, MPC setup with control barrier functions \cite{Wiltz2023, Roque2022, Wabersich2022a, Zeng2021, Brunke2022} as well as those that use robust MPC methods \cite{Bejarano2023, Leeman2023, Fisac2019, Li2020, Wabersich2022}. Here we focus on the discrete-time safety addressed in \cite{Wabersich2022a, Zeng2021, Brunke2022, Bejarano2023, Leeman2023, Fisac2019, Li2020, Wabersich2022}, which unfortunately do not provide hard guarantees of safety for perturbed, nonlinear, time-varying systems with time-varying constraints. In \cite{Wabersich2022a}, a predictive safety filter is developed for  time-invariant, nonlinear systems that is always guaranteed to be feasible and ensures asymptotic stability to the safe set. In \cite{Zeng2021}, safety and asymptotic stability are addressed by combining MPC with control barrier functions for nonlinear, time-invariant systems.  In \cite{Brunke2022}, a disturbance observer is combined with an MPC-based barrier function control law to guarantee the safety of perturbed nonlinear, time-invariant systems with measurement noise. Other types of predictive safety filters have been developed for handling perturbations, but are restricted to time-invariant systems and require global bounds on the disturbance \cite{Bejarano2023}, are restricted to linear time-invariant systems \cite{Leeman2023}, or focus on probabilistic safety \cite{Fisac2019, Li2020, Wabersich2022}. We note that there exist many robust, nonlinear MPC methods, but many time-varying implementations are focused on linear systems \cite{Nguyen2023, Berberich2022a} or adaptive methods \cite{Dhar2021, Bruggemann2022a} wherein there is control over how the the time-varying components update. Here we are interested in the development of a robust, predictive safety filter for time-varying systems with time-varying constraints. This sole focus on safety (instead of safety and performance common to MPC) allows for less restrictive conditions to be satisfied, can yield additional properties such as convergence to the safe region \cite{Wabersich2022a}, and as will be shown here can reduce the amount of computation needed online.  

As mentioned previously, predictive safey encompasses the 1-step safety filters common to barrier function methods. Existing discrete-time barrier function methods have been applied to bipedal robotics \cite{Agrawal2017} and adaptive high-order systems \cite{Xiong2022}, for example, but there is a limited range of methods for handling bounded perturbations. \cite{Takano2018} presents a robust barrier function that uses GP estimate of disturbance, but no hard guarantees of safety are provided. \cite{Cosner2023} presents barrier functions that are robust to stochastic perturbations. \cite{Ahmadi2019} use barrier functions for multi-agent partially observable Markov decision
processes for probabilistic safety guarantees. \cite{Zeng2021a} uses MPC in combination with barrier functions, but no formal guarantees of safety are provided. \cite{Liu2022a} combines high-order barrier functions in an MPC framework. \cite{Zeng2021} combines nonlinear MPC with barrier functions, focusing on enhancing feasibility. \cite{Thirugnanam2022} uses MPC with barrier functions for obstacle avoidance. Apart from \cite{Takano2018}, \cite{Cosner2023} \cite{Wabersich2022a}, and \cite{Brunke2022}, none of the previous methodologies address robustness to perturbations. The methods from \cite{Takano2018} \cite{Cosner2023} only address probabilistic safety, whereas the asymptotic stability results from \cite{Wabersich2022a} are effective for vanishing perturbations, but not for bounded, non-vanishing perturbations which are common in practice. In the continuous-time domain, there exist many solutions that are robust to bounded perturbations \cite{Cohen2022b, Tan2022b}, but this has been lacking in the discrete-time domain. Furthermore, many of the aforementioned approaches require constant online computation of optimization problems, which can be restrictive for computationally limited edge devices. A new safety filter is needed that is robust, addresses time-varying systems/constraints, and is only solved if the nominal control will violate constraints.

In this work, we extend the predictive safety filter \cite{Wabersich2022a} to address bounded perturbations as well as time-varying systems and constraints. Furthermore, we address computation demand via an event-triggering scheme. This event triggering only requires solving the safety filter optimization problem if the nominal control performs an unsafe action. We also extend the existing results in the literature to address robust discrete-time barrier functions with a 1-step safety filter that reduces computation and simplifies the process of barrier function verification.
The resulting methodology is guaranteed to be safe and robust. Figure~\ref{fig:graphical_abstract} illustrates the proposed method. The main contributions of this work are: (1) Robust predictive safety filter for time-varying systems and constraints, (2) 1-step robust safety filter for discrete-time systems, and (3) event-triggering methods for reducing online computations.

The paper is outlined as follows. In Section \ref{sec:background} we introduce the notation, relevant preliminaries, and the problem formulation. In Section \ref{sec:robust predictive safety}, we introduce the robust barrier function formulation and the robust predictive safety filter. In Section \ref{sec:extensions}, we extend the predictive safety filter with an event-triggering scheme and develop a 1-step robust safety filter for robust discrete-time barrier functions. In Section \ref{sec:examples}, we apply the proposed method to a two-tank system, a building system, and a single integrator.

\section{Background}\label{sec:background}

\subsection{Notation and Preliminaries}

Let $\mathbb{N}$ be the set of natural numbers including $0$. The notation $\Lip{f}{k}$ is used to denote the Lipschitz constant of a Lipschitz function, $\myvar{f}(\myvar{x}, k)$, with respect to the second argument. Similarly, $\Lip{f}{x}$ is used to denote the Lipschitz constant of $\myvar{f}$ with respect to the first argument. The notation $\Lip{f}{}$ will be used to denote a bound with respect to the function $\myvar{f}$ and will be explicitly defined in the text.

We say a function $z: \mathbb{R}^n \times \mathbb{N} \to \mathbb{R}$ is locally Lipschitz continuous on $\myset{D} \subset \mathbb{R}^n$ if $z$ satisfies the following Lipschitz condition:
\begin{equation*}
    \|z(\myvar{x}, k) - z(\myvar{y},k)\| \leq \Lip{z}{x} \|\myvar{x} - \myvar{y}\|
\end{equation*}
for all $\myvar{x}, \myvar{y} \in \myset{D}$, $k \in \mathbb{N}$. Unless otherwise stated, $z$ \emph{can} be discontinuous with respect to $k$.

We say a function $z: \mathbb{R}^n \times \mathbb{N} \to \mathbb{R}$ is  1-step bounded on $\mathbb{N}$ with $\Lip{z}{k}$ if $z$ satisfies the following condition:
\begin{equation}\label{eq:one step time bounded}
    \|z(\myvar{x}, k+1) - z(\myvar{x},k)\| \leq  \Lip{z}{k}
\end{equation}
for all $\myvar{x}\in \mathbb{R}^n$, $k \in \mathbb{N}$.

\subsection{Problem Formulation}

Consider the following discrete-time dynamical system:
\begin{equation}\label{eq:nonlinear system discrete}
    \myvar{x}_{k+1} = \myvar{f}(\myvar{x}_k, \myvar{u}_k, k) + \myvar{d}(\myvar{x}_k, \myvar{u}_k, k), \ \myvar{x}_0 \in \mathbb{R}^n, k \in \mathbb{N}
\end{equation}
where $\myvar{f}: \mathbb{R}^n \times \mathbb{R}^m \times \mathbb{N} \to \mathbb{R}^n$ is a function defining the known model, $\myvar{u} \in \myset{U}$ is the control input for the compact set $\myset{U} \subset \mathbb{R}^m$, and $\myvar{d}:\mathbb{R}^n \times \mathbb{R}^m \times \mathbb{N} \to \myset{D} \subset \mathbb{R}^n$ is a function defining the unknown disturbance.

Next, we define the safe region. Consider a function $b: \mathbb{R}^n\times \mathbb{N} \to \mathbb{R}$. This function $b$ is used to define general system constraints that should be satisfied, whose related safe set is defined as follows:
\begin{equation}\label{eq:system constraint set}
    \myset{X}(k) := \left\{\myvar{x} \in \mathbb{R}^n : b(\myvar{x}, k) \geq 0  \right\}
\end{equation}

The problem addressed here is to design a control law that implements a nominal control law, denoted $\myvar{u}_{nom}: \mathbb{R}^n \times \mathbb{N} \to \myset{U}$, while always making sure the system remains in the safe set $\myset{X}(k)$. Here the nominal control could be any previously defined control law or learning-based control that has no safety guarantees associated with it. The purpose of the proposed control is to only override the nominal control to ensure safe action. 
\begin{problem}\label{prob:main}
    Consider the system \eqref{eq:nonlinear system discrete} with unknown disturbance. Given a safe region $\myset{X}(k)$ from \eqref{eq:system constraint set}  defined for all $k \in \mathbb{N}$, and any nominal control $\myvar{u}_{nom}: \mathbb{R}^n \times \mathbb{N} \to \myset{U}$, design a control law $\myvar{u}_k: \mathbb{R}^n \times \mathbb{N} \to \myset{U}$ such that $\myvar{x}_k \in \myset{X}(k)$ for all $k \in \mathbb{N}$. 
\end{problem}

\section{Robust Predictive Safety Filter}\label{sec:robust predictive safety}

\subsection{Robust Barrier Functions}

In this section, we present the proposed predictive safety filter method to address Problem  \ref{prob:main}. The approach considered here is based on control barrier functions for discrete-time systems. 

Let $h: \mathbb{R}^n \times \mathbb{N} \to \mathbb{R}$ denote a barrier function and let the corresponding time-varying set be defined by:
\begin{equation}\label{eq:safe set tv}
    \myset{C}(k) = \{ \myvar{x} \in \mathbb{R}^n: h(\myvar{x},k) \geq 0 \}
\end{equation}

The main idea behind the barrier function is that if $h$ satisfies certain dynamic conditions for \eqref{eq:nonlinear system discrete}, then for any $\myvar{x}_k \in \myset{C}(k)$, for any time $k \in \mathbb{N}$, there exists a control law to ensure $\myset{C}(k)$ is forward invariant, i.e., if $\myvar{x}_0 \in \myset{C}(0)$, then  $\myvar{x}_k \in \myset{C}(k)$ for all $k \in \mathbb{N}$. Thus if $\myset{C}(k) \subset \myset{X}(k)$ for all $k\in \mathbb{N}$ and $\myvar{x}_0 \in \myset{C}(k)$, then it follows that the safety condition from Problem \ref{prob:main} is satisfied: $\myvar{x}_k \in \myset{X}(k)$ for all $k \in \mathbb{N}$. The key then is to determine what dynamic conditions $h$ needs to satisfy in order to guarantee the safety condition in the presence of unknown disturbances. We will also extend this notion using prediction so that the dynamic condition only needs to hold at a horizon $N \in \mathbb{N}$, $N\geq 1$.

Before presenting the dynamic condition to address unknown disturbances we make the following realistic assumptions on \eqref{eq:nonlinear system discrete} and $h$:
\begin{assumption}\label{asm:lipschitz f} The system \eqref{eq:nonlinear system discrete} satisfies the following:
\begin{enumerate}
    \item The function $\myvar{f}: \mathbb{R}^n \times \mathbb{R}^m \times \mathbb{N} \to \mathbb{R}^n $is locally Lipschitz continuous on a set $\myset{W} \subset \mathbb{R}^n$, for which $\myset{W} \supset \myset{X}(k)\  \forall k \in \mathbb{N}$, with Lipschitz constant $\Lip{f}{x} \in \mathbb{R}_{\geq 0}$.
    \item For the unknown disturbance in \eqref{eq:nonlinear system discrete}, there exists a bound $\Lip{d}{} \in \mathbb{R}_{\geq 0}$, $\Lip{d}{} < \infty$,  satisfying:
\begin{align}\label{eq:d bound}
    \Lip{d}{} := \hspace{0.1cm} & \underset{\myvar{x} \in \myset{X}(k), \myvar{u} \in \myset{U}, k\in \mathbb{N} } {\text{sup}}
\hspace{.3cm} || \myvar{d}(\myvar{x}, \myvar{u}, k)||  
\end{align}
\end{enumerate}
\end{assumption}
\begin{assumption}\label{asm:lipschitz}
    For $N \in \mathbb{N}$, $N \geq 1$, let $\myset{Y} \subset \mathbb{R}^n$ satisfy:
        \begin{equation*}
            \myset{Y} \supset \bigcup\limits_{\myvar{x} \in \myset{X}(k), k \in \mathbb{N}} \left\{\myvar{y} \in \mathbb{R}^n: \|\myvar{y}-\myvar{x} \| \leq \Lip{d}{}\sum_{j=0}^{N-1} (\Lip{f}{x})^j\right\}
        \end{equation*}. The following Lipschitz conditions hold:
    \begin{enumerate}
        \item The function $b: \mathbb{R}^n \times \mathbb{N} \to \mathbb{R}$ is locally Lipschitz continuous on $\myset{Y}$
        with Lipschitz constant $\Lip{b}{x} \in \mathbb{R}_{\geq 0}$. 
        \item The function $h: \mathbb{R}^n \times \mathbb{N} \to \mathbb{R}$ is locally Lipschitz continuous on  $\myset{Y} $, with Lipschitz constant $\Lip{h}{x} \in \mathbb{R}_{\geq 0}$. 
    \end{enumerate}
    
\end{assumption}
\begin{remark}
Many dynamical systems satisfy the Lipschitz condition in Assumption \ref{asm:lipschitz f} including robotics and buildings \cite{ShawCortez2022a, Drgona2020}, which is less restrictive than the smoothness requirement in similar MPC settings \cite{Berberich2022a}. Also, the disturbance condition is a common assumption \cite{Bejarano2023} and can be satisfied directly if a) $\myvar{d}$ is continuous on its domain, $\myset{C}(k)$ is compact for all time, and $\myvar{d}$ is bounded in time, or b) $\myset{D}$ is compact. Many existing time-invariant barrier function methods assume $\myset{C}(k)$ is compact which is a special case of this assumption \cite{Wabersich2022a}. Furthermore, Assumption \ref{asm:lipschitz} restricts only how $h$ and $b$ change with respect to $\myvar{x}$. This assumption still allows for discontinuous jumps in time, which is reflected in many building applications. The Lipschitz condition in Assumption \ref{asm:lipschitz} takes into account the effect of the disturbance on $b$ and $h$. The condition says that set $\myset{Y}$ must contain the set $\myset{X}(k)$ in addition to a ball, dependent on the disturbance bound, surrounding each point in $\myset{X}(k)$. Finally, there exist methods to compute Lipschitz constants for general systems \cite{Khalil2002} as well as neural-network models used in system identification \cite{Pauli2022,Fazlyab2019}.
\end{remark}

Now we define the associated safety condition in a \emph{predictive} context. For a horizon $N$, the associated dynamic condition is:
\begin{multline}\label{eq:dcbf predictive}
    \delta \bar{h}_{N}(\myvar{x}_k, \myvar{u}_k, k):= h( \myvar{f}(\myvar{x}_k, \myvar{u}_{k}, k), k+ 1)  -\Lip{h}{x}\Lip{d}{} (\Lip{f}{x})^{N-1} \\ \geq 0
\end{multline}
The condition in \eqref{eq:dcbf predictive} requires that at the end of the $N$-step horizon, the predicted system needs to be sufficiently far into the safe region to take into account the worst possible disturbance. We define the \emph{N-step} \RBF \hspace{0.1mm} as follows:

\begin{definition}\label{def:barrier Nstep}
Consider the system \eqref{eq:nonlinear system discrete} and the function $h: \mathbb{R}^n \times \mathbb{N} \to \mathbb{R}$ with $\myset{C}(k)$ defined by \eqref{eq:safe set tv} and 
\begin{equation}\label{eq:control safe set predictive}
    \bar{\myset{K}}(\myvar{x},k):= \{ \myvar{u} \in \myset{U}: \delta \bar{h}_N(\myvar{x}, \myvar{u}, k) \geq 0, \text{ if } \myvar{x} \in \myset{C}(k)\}.
\end{equation}
We say $h$ is an \emph{N-step} \RBF, if $\bar{\myset{K}}(\myvar{x},k)\neq \emptyset$, $\forall \myvar{x} \in \myset{C}(k)$, $\forall k \in \mathbb{N}$. 
\end{definition}

\begin{remark}
    The motivation behind \eqref{eq:dcbf predictive} is based on recent results for discrete-time barriers \cite{Freire2023}. Consider the conventional time-invariant barrier function where $N =1$ and no disturbance exists such that \eqref{eq:dcbf predictive} becomes: $h(\myvar{x}_{k+1}) \geq 0$. This is a necessary and sufficient condition for ensuring forward invariance of $\myset{C}$ as shown in \cite{Freire2023}. This is notably different from other discrete-time barriers \cite{Xiong2022, Ahmadi2019} which adopt the \emph{continuous}-time formulation: $h(\myvar{x}_{k+1}) - h(\myvar{x}_{k}) \geq -\gamma (h(\myvar{x}_{k}))$ which rearranges to: $h(\myvar{x}_{k+1}) \geq h(\myvar{x}_k)-\gamma (h(\myvar{x}_{k})) $ for an extended class-$\myset{K}$ function $\gamma:\mathbb{R}\to \mathbb{R}$ satisfying $\gamma(r) \leq r$. Although it seems that $\gamma$ allows for freedom in the design of $h$, in the discrete-time case this is not true. Consider the simple choice of a scalar $\gamma$ for which the condition $\gamma(r) \leq r$ becomes $\gamma \leq 1$, the difference condition becomes $h(\myvar{x}_{k+1}) \geq (1 - \gamma ) (h(\myvar{x}_{k}))$.  For any $\gamma < 1$, this requires $h(\myvar{x}_{k+1})$ to be strictly greater than zero which is overly restrictive for ensuring safety. The choice of $\gamma = 1$ of course results in the same condition from \cite{Freire2023}. The proposed predictive form in \eqref{eq:dcbf predictive} extends the recent development from \cite{Freire2023} to account for prediction and robustness in the barrier function.
\end{remark}

The \emph{N-step} \RBF \ will be used to ensure there always exists a control $N$ steps in the future to enforce safety. Before presenting the safety filter, we present a result on how to ensure future state trajectories are safe for an $N$-step look-ahead. To do so, we first define the prediction of a future state as:
\begin{align}\label{eq:future traj}
    &\robvar{x}_{N|k} = \robvar{f}(\myvar{x}_k, \{\myvar{u}_{l|k}\}_{l=0}^{N-1}, k) := \nonumber\\
    & \myvar{f}( ...\myvar{f}(\myvar{x}_k, \myvar{u}_{0|k}, k),  ... \myvar{u}_{N-1|k}, k+N-1 )
\end{align}
where the notation $\{\myvar{u}_{l|k}\}_{l=0}^N:= \{\myvar{u}_0, \myvar{u}_1,...,\myvar{u}_N\}$ is used to denote the set of inputs $\myvar{u}_{l+k}$ starting from time $l=0$ up to $l=N$. The notation $\myvar{x}_{N|k}$ is the state $\myvar{x}_{k+N}$ at time $k+N$ beginning from time $k$ at which point $\robvar{x}_{0|k} = \myvar{x}_k$.

Next, we define the following term to accommodate future predictions with respect to $\myset{X}(k)$:
\begin{multline}\label{eq:predictive robust condition}
    \delta \hat{b}_{N|k}(\myvar{x}_k, \{\myvar{u}_{l|k}\}_{l=0}^{N-1}, k):= b( \robvar{f}(\myvar{x}_k, \{\myvar{u}_{l|k}\}_{l=0}^{N-1}, k), k+ N) \\
    - \Lip{b}{x} \Lip{d}{} \sum_{j=0}^{N-1} (\Lip{f}{x} )^j \geq 0
\end{multline}
Finally, we define the following set of safety-admissible control actions as:
\begin{multline}\label{eq:predictive control set}
    \robset{K}_{N|k}(\myvar{x}_k, k):= \left\{ \myvar{u}_{i-1+k} \in \myset{U}, \ i\in [1,N]\subset \mathbb{N}: \right. \\ 
    \left. \delta \hat{b}_{i|k}(\myvar{x}_k, \{\myvar{u}_{l|k}\}_{l=0}^{i-1}, k) \geq 0 \right\}
\end{multline}
In the following proposition, we show that control inputs that lie inside the predictive set of safety admissible actions from \eqref{eq:predictive control set} ensure open-loop safety of the system over the finite horizon.
\begin{proposition}\label{prop:predictive safety}
    Consider the system \eqref{eq:nonlinear system discrete} for which Assumption \ref{asm:lipschitz f} holds. Given a function $b: \mathbb{R}^n\times  \mathbb{N} \to \mathbb{R}$ satisfying Assumption \ref{asm:lipschitz}.1, let $\robset{K}_{N|k}$, $\delta \hat{b}_{N|k}$, and $\myset{X}(k)$ be defined by \eqref{eq:predictive control set},  \eqref{eq:predictive robust condition}, and \eqref{eq:system constraint set} respectively. For a given $\myvar{x}_{k_0} \in \myset{X}(k_0)$, and $k_0, N \in \mathbb{N}$, $N\geq 1$, if $\{\myvar{u}_{l|k_0}\}_{l=0}^{N-1} \in \robset{K}_{N|k_0}(\myvar{x}_{k_0}, k_0)$ is applied to \eqref{eq:nonlinear system discrete} in open-loop from time $k_0$ to $N-1$, then $\myvar{x}_{k} \in \myset{X}(k)$ for all $k \in [k_0, k_0 + N] \subset \mathbb{N} $. 
\end{proposition}
\begin{proof}
     Let $i \in [1, N] \subset \mathbb{N}$. The proof follows by showing that the error between the unperturbed prediction $\robvar{x}_{i|k}$ (defined by \eqref{eq:future traj}) and perturbed prediction $\myvar{x}_{i|k}$ (defined by \eqref{eq:nonlinear system discrete}) is appropriately bounded and that this bound is taken into account via $\delta \hat{b}_{i|k}$. By definition of $\delta \hat{b}_{N_k}$ and $\robset{K}_{N|k}$, $\robvar{x}_{i|k} \in \myset{X}(k+i) \subset \myset{W}$ for all $i \in [1,N]$, i.e., the future trajectory of the unperturbed system is safe, the dynamics are locally Lipschitz continuous on this trajectory, and the bound on the disturbance holds. This cannot yet be said for for the unperturbed trajectory. 
     
     We can however claim that for some $i \in [1,N]\subset \mathbb{N}$, $\myvar{x}_{i|k}\in \myset{X}(k+i)$. This proof is straightforward by simply choosing $i=1$ and noting that $\robvar{x}_{0|k_0} = \myvar{x}_{0|k_0} = \myvar{x}_{k_0} \in \myset{X}(k_0) \subset \myset{W}$ so the locally Lipschitz property and disturbance bound of Assumption \ref{asm:lipschitz f} hold for $\myvar{x}_{k_0}$, which along with the triangle inequality yields:
     $ \|\robvar{x}_{1|k_0} - \myvar{x}_{1|k_0} \| 
            = \|\myvar{f}(\robvar{x}_{0|k_0}, \myvar{u}_{0|k_0}, k_0) 
            - \myvar{f}(\myvar{x}_{0|k_0}, \myvar{u}_{0|k_0}, k_0) 
             \quad - \myvar{d} (\myvar{x}_{0|k_0}, \myvar{u}_{0|k_0}, k_0) \| 
             \leq \Lip{d}{} + \Lip{f}{x}\|\robvar{x}_{0|k_0} - \myvar{x}_{0|k_0} \|$. 
      From Assumption \ref{asm:lipschitz}, the local Lipschitz property of $b$ applies to $\myvar{x}_{1|k_0}$ and $\robvar{x}_{1|k_0}$, such that the following holds: $|b(\robvar{x}_{1|k_0}, k_0+1) - b(\myvar{x}_{1|k_0}, k_0+1) |  \leq \Lip{b}{x} \|\robvar{x}_{1|k_0} - \myvar{x}_{1|k_0} \|$. Thus from the previous relations we have: $|b(\robvar{x}_{1|k_0}, k_0+1) - b(\myvar{x}_{1|k_0}, k_0+1) | \leq \Lip{b}{x} \Lip{d}{}$. This implies that: $
    b(\myvar{x}_{1|k_0}, k_0+1) \geq b(\robvar{x}_{1|k_0}, k_0+1) - \Lip{b}{x} \Lip{d}{}$.
By assumption, since $\{\myvar{u}_{l|k_0} \}_{l=0}^{N-1} \in \robset{K}_{N|k}(\myvar{x}_{k_0}, k_0)$, $\myvar{u}_{0|k_0}$ satisfies: $\delta \hat{b}_{1|k_0}(\myvar{x}_{k_0}, \{\myvar{u}_{0|k_0} \}, k_0) \geq 0$ for which $b(\robvar{x}_{1|k_0}, k_0+1) \geq   \Lip{b}{x} \Lip{d}{} $. This yields: $b(\myvar{x}_{1|k_0}, k_0+1) \geq b(\robvar{x}_{1|k_0}, k_0+1) - \Lip{b}{x} \Lip{d}{} \geq 0$ such that $\myvar{x}_{1|k_0} = \myvar{x}_{k_0+1} \in \myset{X}(k_0+1)$.
       
To complete the proof, we repeat the above argument recursively for $i = 2,...,N$. Note that for a given $i$, $\myvar{x}_{i-1|k_0} \in \myset{X}(k_0+i-1)$ so that the Lipschitz property of $f$ always holds. Next we define the following relations for any $i \in {1,...,N}$. Using the Lipschitz property of $\myvar{f}$ and the bound $\Lip{d}{}$ along with the triangle inequality yields: $ \|\robvar{x}_{i|k_0} - \myvar{x}_{i|k_0} \|  = \|\myvar{f}(\robvar{x}_{i-1|k_0}, \myvar{u}_{i-1|k_0}, k_0+i-1) 
            - \myvar{f}(\myvar{x}_{i-1|k_0}, \myvar{u}_{i-1|k_0}, k_0+i-1) 
             \quad - \myvar{d} (\myvar{x}_{i-1|k_0}, \myvar{u}_{i-1|k_0}, k_0+i-1) \| 
             \leq \Lip{d}{} + \Lip{f}{x}\|\robvar{x}_{i-1|k_0} - \myvar{x}_{i-1|k_0} \|$.
    We repeatedly apply the above process until we reach $i =1$. This can be written as follows for $F(y) = \Lip{d}{} + \Lip{f}{x} y$ and $F_i(y) = \underbrace{F\circ F\circ ... \circ F}_{i \text{ times}}(y)$: $\|\robvar{x}_{i|k} - \myvar{x}_{i|k} \| \leq \underbrace{F\circ F\circ ...\circ F}_{i \text{ times}} (\|\robvar{x}_{0|k} - \myvar{x}_{0|k} \|) 
         = F_i(\|\robvar{x}_{0|k} - \myvar{x}_{0|k} \| )$.
Now since $\robvar{x}_{0|k_0} = \myvar{x}_{0|k_0} = \myvar{x}_{k_0}$, it is clear from the previous relation that $\|\robvar{x}_{i|k} - \myvar{x}_{i|k} \| \leq F_i(0)$ for which by evaluation $F_i(0) = \Lip{d}{} \sum_{j=0}^{i-1} (\Lip{f}{x})^j$.

Now it is clear that from the bound defined by $F_i(0)$, the Lipschitz property of $b$ applies to each $\myvar{x}_{i|k_0}$ so that the following holds: $|b(\robvar{x}_{i|k_0}, k_0+i) - b(\myvar{x}_{i|k_0}, k_0+i) |  \leq \Lip{b}{x} \|\robvar{x}_{i|k_0} - \myvar{x}_{i|k_0} \|$. Thus from the previous relations we have: $|b(\robvar{x}_{i|k_0}, k_0+i) - b(\myvar{x}_{i|k_0}, k_0+i) | \leq \Lip{b}{x} \Lip{d}{} \sum_{j=0}^{i-1} (\Lip{f}{x})^j$. This implies that:
\begin{equation}\label{eq:prediction h relation}
    b(\myvar{x}_{i|k_0}, k_0+i) \geq b(\robvar{x}_{i|k_0}, k_0+i) - \Lip{b}{x} \Lip{d}{} \sum_{j=0}^{i-1} (\Lip{f}{x})^j
\end{equation}
By assumption, since $\{\myvar{u}_{l|k_0} \}_{l=0}^{N-1} \in \robset{K}_{N|k}(\myvar{x}_{k_0}, k_0)$, $\{\myvar{u}_{l|k_0} \}_{l=0}^{i-1}$ satisfies: $\delta \hat{b}_{i|k_0}(\myvar{x}_{k_0}, \{\myvar{u}_{l|k_0} \}_{l=0}^{i-1}, k_0) \geq 0$ for which $b(\robvar{x}_{i|k_0}, k_0+i) \geq   \Lip{b}{x} \Lip{d}{} \sum_{j=0}^{i-1} (\Lip{f}{x})^j$ (see \eqref{eq:future traj}, \eqref{eq:predictive robust condition}). Combined with \eqref{eq:prediction h relation} yields: $b(\myvar{x}_{i|k_0}, k_0+i) \geq b(\robvar{x}_{i|k_0}, k_0+i) - \Lip{b}{x} \Lip{d}{} \sum_{j=0}^{i-1} (\Lip{f}{x})^j \geq 0$ such that $\myvar{x}_{i|k_0} = \myvar{x}_{k_0+i} \in \myset{X}(k_0+i)$. Application of this for all $i \in [1, N] \subset \mathbb{N}$ with $k = k_0+i$ completes the proof.
\end{proof}

It is important to emphasize that Proposition \ref{prop:predictive safety} does \emph{not} require that $b$ is a barrier function in any sense. This result is a sufficient condition for ensuring that in the finite horizon, the safety conditions will be met. The $\emph{N-step}$ \RBF \ will then ensure that the open-loop control from Proposition \ref{prop:predictive safety} can be applied for all future time to enforce safety. This will be presented in the next subsection.

\subsection{Predictive Control}

Here we present the robust predictive safety filter control. Consider $h$ as an \emph{N-step} \RBF. The robust predictive safety filter is defined by the following \textbf{safety filter problem}, respectively:
\begin{subequations}\label{eq:psf original}
\begin{flalign}
    & \myset{P}_{sf}(\myvar{x}_k, k)= \nonumber &&\\
  \vspace{0.1cm}  &  \hspace{0.5cm}  \underset{\myvar{u}_{l|k} } {\text{argmin}} \hspace{.3cm} \|\myvar{u}_{nom}(\myvar{x}_k, k) - \myvar{u}_{0|k}\|  &&\\
  & \hspace{0.5cm} \text{s.t. }   \forall l = 0,...,N-1: \nonumber&&\\
& \hspace{0.5cm}  \robvar{x}_{0|k} = \myvar{x}_k,  &&\\
& \hspace{0.5cm}  \robvar{x}_{l+1|k} = \myvar{f}(\robvar{x}_{l|k}, \myvar{u}_{l|k},l+k),  \label{eq:dynamics unperturbed} &&\\
& \hspace{0.5cm}  \myvar{u}_{l|k} \in \myset{U}, &&\\
& \hspace{0.5cm}  b(\robvar{x}_{l|k}, l+k) \geq    \Lip{b}{x} \Lip{d}{} \sum_{j=0}^{l-1} {(\Lip{f}{x})}^{j} , \label{eq:psf original state}&&\\
& \hspace{0.5cm}  h(\robvar{x}_{N|k}, k+N) \geq   \Lip{h}{x}\Lip{d}{} (\Lip{f}{x})^{N-1}  \label{eq:psf original terminal} &&
\end{flalign}
\end{subequations}
\begin{equation}\label{eq:safety filter control}
    \{\myvar{u}^*_{l|k}\}_{l=0}^N  = \myset{P}_{sf}(\myvar{x}_k, k)
\end{equation}

\begin{remark}
There are several differences between the proposed robust predictive safety filter and that of \cite{Wabersich2022a}. First, the proposed safety filter only requires the solution of one optimization problem whereas \cite{Wabersich2022a} requires an additional feasibility problem to be solved. Second, the  robustness-related terms, i.e., $\Lip{b}{x} , \Lip{d}{}$, $\Lip{h}{x}$, and $\Lip{f}{x}$ ensure any bounded disturbance cannot push the system outside of the safe set, which is not ensured in \cite{Wabersich2022a}.
\end{remark}

The barrier function $h$ acts as a terminal constraint and must be an \emph{N-step} \RBF \hspace{0.1mm} for robustness guarantees to hold. The $h$ function must be constructed to satisfy the conditions of Definition \ref{def:barrier Nstep}, which will most likely result in a conservative terminal set. Thus for more aggressive, but safe performance, a designer would construct $h$ conservatively, but then enlarge the prediction horizon, $N$. The constraints defined by $b$ can then be satisfied without checking that $b$ itself is a barrier function in any sense. The trade-off is that as $N$ increases, more computation is required to check that safety is guaranteed, while the system state is able to deviate further from the conservative bound defined by $h$. For a time-invariant $h$, the linearization approach from \cite{Wabersich2022a} can be used to construct  $h$ to satisfy Definition \ref{def:barrier Nstep}.

\subsection{Analysis}

The following assumption is made regarding the terminal constraint function $h$:
 \begin{assumption}\label{asm: h is rbf}
     The function $h: \mathbb{R}^n \times\mathbb{N} \to \mathbb{R}$  with corresponding sets $\myset{C}(k)$ and $\bar{\myset{K}}$ defined by \eqref{eq:safe set tv} and \eqref{eq:control safe set predictive}, respectively, is an \emph{N-step} \RBF. 
 \end{assumption}

In the following theorem, we show that the proposed control is always feasible and ensures the states remain inside of the safe set $\myset{X}(k)$ for all time in the presence of disturbances.  Let the robust safety set and robust terminal safe set be respectively defined by:
\begin{multline}\label{eq:robust safe set}
    \myset{X}^l(k) = \{\myvar{x} \in \mathbb{R}^n: b(\myvar{x}, k) \geq  \Lip{b}{x} \Lip{d}{} \sum_{j=0}^{l-1} {(\Lip{f}{x})}^{j} \}, \\
    \forall l \in \{0, ..., N\}
\end{multline}
\begin{equation}\label{eq:robust safe terminal set}
    \myset{X}_{f}^N(k) = \{\myvar{x} \in \mathbb{R}^n: h(\myvar{x}, k) \geq   \Lip{h}{x}\Lip{d}{} (\Lip{f}{x})^{N-1}  \}
\end{equation}
The following assumption is required to ensure the robust safety sets are non-empty and that if the final state in the horizon lies in terminal robust safety set, then it also lies in the robust safety set, i.e.,  $\myset{X}_{f}^N(k+N) \subset \myset{X}^N(k+N), \ \forall k \in \mathbb{N}$. 
\begin{assumption}\label{asm:nonempty sets}
    The following conditions hold:
    \begin{enumerate}
        \item $\myset{X}^l(k+l) \neq \emptyset$, $\forall l \in \{0,...,N\}, \ \forall k \in \mathbb{N}$
        \item $\myset{X}_{f}^N(k) \neq\emptyset$, $\forall k \in \mathbb{N}$ 
        \item $\myset{X}_{f}^N (k+N) \subset \myset{X}^{N}(k+N)  $, $\forall k \in \mathbb{N}$.
    \end{enumerate}
\end{assumption}
\begin{remark}
    The last condition in Assumption \ref{asm:nonempty sets} requires that the  robust safety region at prediction time $N$ must contain the level set defined by the \emph{N-step} \RBF. The idea here is that we treat $\myset{X}_f^N$ as the terminal set, and so if the system reaches this terminal set at time $N$, then it will satisfy the system constraints defined by $\myset{X}$ while taking into account the effects of the perturbation. This condition can be satisfied by shrinking $\myset{C}(k)$ appropriately.
\end{remark}

The following theorem presents one of the main results and states that the proposed robust safety filter ensures the states remain inside of the safe region $\myset{X}(k)$ and feasibility of the proposed control for all time. The main differences between this result and that of \cite{Wabersich2022a} is that here we address robustness to perturbations and extend the approach to time-varying systems with time-varying constraints. The key in this proof is in handling the disturbances in all future predictions of the system by exploiting Proposition \ref{prop:predictive safety} and Definition \ref{def:barrier Nstep}.
 \begin{thm}\label{theorem:predictive safety}
     Consider the system \eqref{eq:nonlinear system discrete} satisfying Assumption \ref{asm:lipschitz f} and suppose a nominal control law $\myvar{u}_{nom}: \mathbb{R}^n \times \mathbb{N} \to \myset{U}$ is given. Given functions $b: \mathbb{R}^n\times \mathbb{N} \to \mathbb{R}$ and $h: \mathbb{R}^n\times \mathbb{N} \to \mathbb{R}$ for which Assumption \ref{asm:lipschitz} holds, let $\robset{K}$, $\delta \hat{b}_{N|k}$, and $\myset{X}(k)$ be defined by \eqref{eq:predictive control set}, \eqref{eq:predictive robust condition}, and \eqref{eq:system constraint set}, respectively. Suppose Assumptions \ref{asm: h is rbf} and \ref{asm:nonempty sets} hold. For \eqref{eq:nonlinear system discrete} in closed-loop with \eqref{eq:psf original}, \eqref{eq:safety filter control}, and for any given $k_0 \in \mathbb{N}$, $\myvar{x}_{k_0} \in \myset{X}(k_0)$ for which $\myset{P}_{sf}(\myvar{x}_{k_0}, k_0)$ is feasible, the following statements hold:
     \begin{enumerate}
        \item The control defined by \eqref{eq:psf original}, \eqref{eq:safety filter control} is feasible for all $k \geq k_0$.
         \item $\myvar{x}_{k} \in \myset{X}(k)$ for all $k\geq k_0, k \in \mathbb{N}$.
     \end{enumerate}
 \end{thm}
 \begin{proof}
     
 Proposition \ref{prop:predictive safety} ensures that the open-loop application of $\{\myvar{u}^*_{l|k}\}_{l=0}^{N-1}$ to \eqref{eq:nonlinear system discrete} is safe, i.e., $\myvar{u}_{k} = \myvar{u}^*_{0|k}, \myvar{u}_{k+1} = \myvar{u}^*_{1|k}, ..., \myvar{u}_{k+N-1} = \myvar{u}^*_{N-1|k}$, ensures that $\myvar{x}_{\kappa} \in \myset{X}(\kappa)$ for all $\kappa \in \{k,...,k+N\}$. Thus since $\myset{P}_{sf}(\myvar{x}_{k_0}, k_0)$ is feasible, we thus know that $\myvar{x}_{k_0+1} \in \myset{X}(k_0+1)$.

In the closed-loop response, we want to show that $\myset{P}_{sf}$ is always feasible so that we can repeatedly apply Proposition \ref{prop:predictive safety} to ensure safety holds for all future time. In order to show that feasibility of $\myset{P}_{sf}(\myvar{x}_{k}, k)$ implies feasibility of $\myset{P}_{sf}(\myvar{x}_{k+1}, k+1) $, we need to show that there exists a feasible control action at $k+1$.

We will denote the resulting state trajectory associated with $\{\myvar{u}^*_{l|k}\}_{l=0}^{N-1}$ applied to the unperturbed system \eqref{eq:dynamics unperturbed} as $\{\robvar{x}_{l|k} \}_{l=0}^{N}$, i.e., this is the predicted system trajectory with the optimal control at time $k$.
The feasible control action chosen is defined as follows:
\begin{equation}\label{eq:feasible control}
    \myvar{u}^+_{l|k+1} \in \begin{cases}
        \{ \myvar{u}^*_{l+1|k} \}, \forall l \in \{0,...,N-2\}, \\
        \bar{\myset{K}}(\robvar{x}_{N-1|k+1}, k+N), \text{ for } l=N-1 \\ \text{ \hphantom{$\bar{\myset{K}}(\robvar{x}_{N-1|k+1}, $} if }  \robvar{x}_{N-1|k+1} \in \myset{C}(k+N), \\
        \{\myvar{0} \}, \text{ for } l=N-1 \text{ if }  \robvar{x}_{N-1|k+1} \notin \myset{C}(k+N)
    \end{cases} 
\end{equation}
for which the chosen control action yields an unperturbed trajectory $\{ \robvar{x}_{l|k+1} \}_{l=0}^{N-1}  $, where $\robvar{x}_{l|k+1}$ is defined by \eqref{eq:dynamics unperturbed} starting at time $k+1$ with $\robvar{x}_{0|k+1} = \myvar{x}_{k+1}$. Note that $\{ \robvar{x}_{l|k+1}\}_{l=0}^{N-1}$ may not be equal to $\{\robvar{x}_{l|k}\}_{l=1}^{N}$ because $\robvar{x}_{0|k+1} = \myvar{x}_{k+1}$, i.e., the initial condition in the prediction of the unperturbed state at time $k+1$ is the result of the \emph{true} system dynamics. 

The last step here is to show that the control action \eqref{eq:feasible control} yields $\robvar{x}_{l|k+1}$ that satisfy $b(\robvar{x}_{l|k+1} , k+l+1) \geq   \Lip{b}{x} \Lip{d}{} \sum_{j=0}^{l-1} (\Lip{f}{x})^j$ for all $l \in \{0,...,N-1\}$ and $h(\robvar{x}_{N|k+1}, k+N+1) \geq \Lip{h}{x}\Lip{d}{} (\Lip{f}{x})^{N-1} $.

We first follow a similar approach to the proof of Proposition \ref{prop:predictive safety} and define the error between the unperturbed trajectory at $k$, $\robvar{x}_{i+1|k}$, and the resulting trajectory from $\myvar{u}^+_{i|k+1}$, $\robvar{x}_{i|k+1}$, for all $i \in \{0,...,N-1\}$ using the Lipschitz property of $\myvar{f}$. Consider the following relation for $i\in \{1,...,N-1\}$:
\begin{align}\label{eq:error in predictions}
    \|\robvar{x}_{i|k+1} - \robvar{x}_{i+1|k} \|  =& \| \myvar{f}(\robvar{x}_{i-1|k+1}, \myvar{u}^+_{i-1|k+1}, k+i) \nonumber\\
    & - \myvar{f}(\robvar{x}_{i|k}, \myvar{u}^+_{i-1|k+1}, k+i) \| \nonumber\\
   \leq  & \Lip{f}{x} \| \robvar{x}_{i-1|k+1} - \robvar{x}_{i|k} \|
\end{align}
For each $i \in \{1,...,N-1\}$ we can repeat the application of \eqref{eq:error in predictions} until $i = 1$, which yields: $\|\robvar{x}_{i|k+1} - \robvar{x}_{i+1|k} \|  \leq (\Lip{f}{x})^i \| \robvar{x}_{0|k+1} - \robvar{x}_{1|k} \|$.
Now recall that $\robvar{x}_{0|k+1} = \myvar{x}_{k+1} = \myvar{f}(\myvar{x}_k, \myvar{u}^*_{0|k}, k) +\myvar{d}(\myvar{x}_k, \myvar{u}^*_{0|k}, k) $ and $\robvar{x}_{1|k} = \myvar{f}(\myvar{x}_k, \myvar{u}^*_{0|k}, k) $. Substitution into the above inequality along with \eqref{eq:d bound} yields: $ \|\robvar{x}_{i|k+1} - \robvar{x}_{i+1|k} \| \leq  (\Lip{f}{x})^i \| \robvar{x}_{0|k+1} - \robvar{x}_{1|k} \| 
    \leq  (\Lip{f}{x})^i \|\myvar{f}(\myvar{x}_k, \myvar{u}^*_{0|k}, k) +\myvar{d}(\myvar{x}_k, \myvar{u}^*_{0|k}, k) 
      -  \myvar{f}(\myvar{x}_k, \myvar{u}^*_{0|k}, k) \| 
    \leq  (\Lip{f}{x})^i \Lip{d}{}$.
Applying the Lipschitz property of $b$ yields:
$| b(\robvar{x}_{i|k+1}, k+i+1) - b( \robvar{x}_{i+1|k}, k+i+1)| \leq \Lip{b}{x}\Lip{d}{} (\Lip{f}{x})^i $, which implies that, for all $i\in \{0,...,N-1\}$:
\begin{equation}\label{eq:translated relationship}
    b(\robvar{x}_{i|k+1}, k+i+1) \geq b( \robvar{x}_{i+1|k}, k+i+1) - \Lip{b}{x}\Lip{d}{} (\Lip{f}{x})^i 
\end{equation}

Recall that if $\myset{P}_{sf}(\myvar{x}_k, k) $ is feasible, for $i \in \{0,...,N-2\}$ $b( \robvar{x}_{i+1|k}, k+i+1) \geq   \Lip{b}{x} \Lip{d}{} \sum_{j=0}^{i} {(\Lip{f}{x})}^{j}$, which yields: $b(\robvar{x}_{i|k+1}, k+i+1) \geq b( \robvar{x}_{i+1|k}, k+i+1) - \Lip{b}{x}\Lip{d}{} (\Lip{f}{x})^i 
    \geq  \Lip{b}{x} \Lip{d}{} \sum_{j=0}^{i} {(\Lip{f}{x})}^{j} 
     - \Lip{b}{x}\Lip{d}{} (\Lip{f}{x})^i 
    =  \Lip{b}{x} \Lip{d}{} \sum_{j=0}^{i-1} {(\Lip{f}{x})}^{j} $.
Thus the chosen $\{\myvar{u}^+_{l|k+1}\}_{l=0}^{N-3}$ and resulting $\{\robvar{x}_{l|k+1}\}_{l=0}^{N-2}$ satisfies \eqref{eq:psf original state} for $l \in \{0,..., N-2\}$. 

To satisfy \eqref{eq:psf original state} for $l = N-1$, we note that since $\myset{X}_{f}^N(k+N) \subset \myset{X}^{N}(k+N)  $ from Assumption \ref{asm:nonempty sets}, $h(\robvar{x}_{N|k}, k+N)$ $ \geq$ $ \Lip{h}{x}\Lip{d}{} (\Lip{f}{x})^{N-1}  $ $\implies $ $b(\robvar{x}_{N|k}, k+N) \geq  \Lip{b}{x} \Lip{d}{} \sum_{j=0}^{N-1} {(\Lip{f}{x})}^{j} $. Substitution into \eqref{eq:translated relationship} for $i = N-1$ yields: $b(\robvar{x}_{N-1|k+1}, k+N) \geq  b(\robvar{x}_{N|k}, k+N) - \Lip{b}{x}\Lip{d}{}(\Lip{f}{x})^{N-1} 
    \geq    \Lip{b}{x} \Lip{d}{} \sum_{j=0}^{N-1} {(\Lip{f}{x})}^{j} 
     - $ $\Lip{b}{x}\Lip{d}{}(\Lip{f}{x})^{N-1} 
    $= $    \Lip{b}{x} \Lip{d}{} \sum_{j=0}^{N-2} {(\Lip{f}{x})}^{j} $.
Thus the chosen $\myvar{u}^+_{N-2|k+1}$ and resulting $\robvar{x}_{N-1|k+1}$  satisfies \eqref{eq:psf original state} for $l = N-1$.

To satisfy \eqref{eq:psf original terminal}, we substitute $h$ with associated Lipschitz constant $\Lip{h}{x}$ (note the same analysis applies to $h$) into \eqref{eq:translated relationship} for $i = N-1$, which yields: $h(\robvar{x}_{N-1|k+1}, k+N) \geq  h( \robvar{x}_{N|k}, k+N) - \Lip{h}{x}\Lip{d}{} (\Lip{f}{x})^{N-1}  
 \geq    \Lip{h}{x}\Lip{d}{} (\Lip{f}{x})^{N-1}  - \Lip{h}{x}\Lip{d}{} (\Lip{f}{x})^{N-1} 
 \geq  0 $.
Thus $\robvar{x}_{N-1|k+1} \in \myset{C}(k+N)$ and $\bar{\myset{K}}(\robvar{x}_{N-1|k+1}, k+N) \neq \emptyset$ since $h$ is an \emph{N-step} \RBF. Thus from \eqref{eq:feasible control}, $\myvar{u}^+_{N-1|k+1} \in \bar{\myset{K}}(\robvar{x}_{N-1|k+1}, k+N)$. By definition, this implies that $h(\robvar{x}_{N|k+1}, k+N+1) = h(\myvar{f}(\robvar{x}_{N-1|k+1}, \myvar{u}^+_{N-1|k+1}, k+N), k+N+1) \geq \Lip{h}{x}\Lip{d}{} (\Lip{f}{x})^{N-1}$. Thus the chosen control $\myvar{u}^+_{N-1|k+1}$ and resulting $\robvar{x}_{N|k+1}$ satisfies \eqref{eq:psf original terminal}.

We have thus constructed a feasible solution of $\{\myvar{u}^+_{l|k+1}\}_{l=0}^{N-1}$ for which all the constraints of $\myset{P}_{sf}(\myvar{x}_{k+1}, k+1)$ are satisfied. This implies that at the next time step, there exists a control such that $\myset{P}_{sf}$ is feasible and $\myvar{x}_{k+1} \in \myset{X}(k+1)$. Since $\myset{P}_{sf}(\myvar{x}_{k_0}, k_0)$ is feasible, this process can be repeated by induction for all $k\geq k_0$ such that the control is feasible for all $k \geq k_0$ and $\myvar{x}_k \in \myset{X}(k)$ for all $k\geq k_0$,  which concludes the proof.
\end{proof}
Theorem \ref{theorem:predictive safety} ensures safety  for all time. There is no explicit restriction on the size of $N$ in order to ensure safety. Adjusting $N$ simply provides a trade-off between computational complexity and conservatism. Smaller values of $N$ result in smaller optimization problems to be solved and smaller robustness margins, however this will yield a more conservative control law since the system state will be forced to remain close to the robust terminal safe set. On the other hand, larger values of $N$ result in larger optimization problems, but allow for states to deviate from the robust terminal safe set so long as at time $N$, the system predictions can be returned to the robust terminal safe set. Also, note that as $N$ increases the robustness margins increase, which may violate Assumption \ref{asm:nonempty sets}.

    
\begin{remark}[Multiple constraints]\label{rem:multiple constraints}
    Note that in $\myset{P}_{sf}$, only one constraint function, $b$, is enforced throughout the trajectory for readability. In practice, multiple $b_i: \mathbb{R}^n \times \mathbb{N} \to \mathbb{R}$ could be implemented in $\myset{P}_{sf}$ so long as each $b_i$ satisfies the same requirements as $b$. To be clear, the following sets would be used in place of $\myset{X}(k)$ and $\myset{X}^l(k)$ for $i \in \{1,...,M\}$: $\myset{X}_i(k) := \{\myvar{x} \in \mathbb{R}^n: b_i(\myvar{x}, k) \geq 0 \}$, $\myset{X}(k) = \cap_{i = 1}^M \myset{X}_i(k)$, $\myset{X}^l_j(k):= \{\myvar{x} \in \mathbb{R}^n: b_i(\myvar{x}, k) \geq  \Lip{b}{x} \Lip{d}{}\sum_{j=0}^{l-1} (\Lip{f}{x})^{j} \}$, $\myset{X}^l( k) = \cap_{i = 1}^M \myset{X}^l_i(k)$.
\end{remark}

\section{Extensions and Implementation}\label{sec:extensions}

Here we make modifications to the robust predictive safety filter to reduce online computation. We first extend the robust predictive safety filter using an even-triggered scheme. We then use the results of the predictive safety filter to provide an efficient 1-step safety filter for existing barrier function implementations.

\subsection{Event-triggered control}

We propose an event-triggered control policy that implements the robust predictive safety filter as long as the system is safe and implements the predictive control policy \eqref{eq:safety filter control} otherwise to ensure safety. This control policy is defined in Algorithm \ref{alg:event-triggered control}.

\begin{algorithm}[!htbp]
\caption{Event-Triggered, Robust Safety Filter}\label{alg:event-triggered control}
\begin{algorithmic}[1]
\State Given: $\myvar{x}_k$, $k$.
\State Initialize $\myvar{x}_{0|k} = \myvar{x}_k$.
\For{$l \in \{0,..., N-1\}$}
\State Compute $\myvar{u}_{{nom}_{l|k}} = \myvar{u}_{nom}(\myvar{x}_{l|k}, k+l)$.
\State Compute $\robvar{x}_{l+1|k} = \myvar{f}(\robvar{x}_{l|k}, \myvar{u}_{nom_{l|k}}, k+l)$.
\EndFor
\State Define $\{\myvar{u}_{nom_{l|k}}\}_{l=0}^{N-1}$.
\State Define the roll-out $\{ \robvar{x}_{l|k}\}_{l=0}^N$ .
\If{ $\robvar{x}_{l|k} \in \myset{X}^l(k)$ for $l \in \{0,...,N-1\}$ and $\robvar{x}_{N|k} \in \myset{X}_f^N(k)$}
    \State \Return $ \myvar{u}_{nom_{0|k}}$.
\Else
    \State Solve \eqref{eq:safety filter control}  for  $\{\myvar{u}^*_{l|k} \}_{l=0}^{N-1}$.
    \State \Return $ \myvar{u}_{0|k}^*$.
\EndIf
\end{algorithmic}
\end{algorithm}

\begin{cor}\label{theorem:safety event trigger Nstep}
    Suppose the conditions of Theorem \ref{theorem:predictive safety} hold. Given a nominal control law $\myvar{u}_{nom}: \mathbb{R}^n\times \mathbb{N}\to \myset{U}$, suppose the system \eqref{eq:nonlinear system discrete} is in closed-loop with the control from Algorithm \ref{alg:event-triggered control}. If $\myset{P}_{sf}(\myvar{x}_{k_0} , k_0)$ is feasible for any $\myvar{x}_{k_0} \in \myset{X}(k_0)$,  $k_0\in \mathbb{N}$, then  $\myvar{x}_{k} \in \myset{X}(k)\ \forall k\geq k_0, k \in \mathbb{N}$.
\end{cor}
\begin{proof}
Follows directly from the proof of Theorem \ref{theorem:predictive safety}.
\end{proof}
\begin{remark}
    The predictive safety filter is solely focused on enforcing safety. This allows for more modularity in its use with existing control laws over MPC methods that solve safety and performance in one control law. By using prediction only for safety, the event-triggering approach in Algorithm \ref{alg:event-triggered control} is straightforward and can be used on compute-limited systems. This is advantageous when coupled with learning-based controllers which require low onboard computational resources to implement, but do not guarantee safety for general systems. 
\end{remark}

 \subsection{Efficient 1-step Robust Safety Filter}

The robust predictive safety filter is dependent on the existence of an \emph{N-step} \RBF, which can be difficult to synthesize or verify. In special cases, we can check for the \RBF \ condition without checking $\robset{K}$ over all $\myset{C}(k)$. For such cases we restrict our attention to $1$-step safety filters typically seen in the literature. Here we show that under stronger conditions, we can simplify the event-triggering conditions and the design of $h$.

Let $\myset{A}(k) \subset \mathbb{R}^n$ be a set encompassing the boundary of $\myset{C}(k)$, defined by $\mysetbound{C}(k) = \{ \myvar{x} \in \mathbb{R}^n: h(\myvar{x}, k) = 0\}$, which is defined by:
\begin{equation}\label{eq:annulus set tv}
    \myset{A}(k) = \{ \myvar{x} \in \mathbb{R}^n: h(\myvar{x},k) \in [0, a] \}
\end{equation}
for some $a \in \mathbb{R}_{>0}$. The set $\myset{A}(k)$ defines the region around the controlled safe set boundary, which is the only region where we care about \emph{enforcing} the safe set condition, i.e., $h \geq 0$. In this respect, we define the set of admissible control inputs for which $h\geq 0$ holds if $\myvar{x} \in \myset{A}(k)$:
\begin{equation}\label{eq:control safe set with A}
    \myset{K}(\myvar{x},k):= \{ \myvar{u} \in \myset{U}: \delta \bar{h}_{N=1}(\myvar{x}, \myvar{u}, k) \geq 0, \text{ if } \myvar{x} \in \myset{A}(k)\}
\end{equation}
We note that in this case, for any $k$ and $\myvar{x} \in \myset{C}(k) \setminus \myset{A}(k)$, then $\myset{K}(\myvar{x}, k) = \myset{U}$ and so we need only check $\myset{A}(k)$ to determine if $h$ satisfies the conditions of Definition \ref{def:barrier Nstep} for $N=1$.

First, in the next theorem we ensure that for a \emph{1-step} \RBF, forward invariance holds for any $\myvar{u} \in \robset{K}$.
\begin{thm}\label{theorem:safety}
Consider the system \eqref{eq:nonlinear system discrete} satisfying Assumption \ref{asm:lipschitz f} and let $N=1$. Given functions $b: \mathbb{R}^n\times \mathbb{N} \to \mathbb{R}$ and $h: \mathbb{R}^n\times \mathbb{N} \to \mathbb{R}$, let $\robset{K}$ and $\myset{X}(k)$ be defined by \eqref{eq:predictive control set} and \eqref{eq:system constraint set}, respectively. Suppose Assumptions \ref{asm:lipschitz}.2 and \ref{asm: h is rbf} hold and $\myset{C}(k) \subset \myset{X}(k)\ \forall k\in \mathbb{N}$. If $\myvar{x}_{k_0} \in \myset{C}(k_0) \subset \myset{X}(k_0)$ for some $k_0 \in \mathbb{N}$ and $\myvar{u}_k \in \robset{K}(\myvar{x}_k, k)$ is implemented in closed-loop with \eqref{eq:nonlinear system discrete}, then $\myvar{x}_{k} \in \myset{C}(k) \subset \myset{X}(k)$ for all $k \geq k_0, k \in \mathbb{N}$.
\end{thm}
\begin{proof}
Due to the restriction to $N=1$, this proof is slightly different to that of Theorem \ref{theorem:predictive safety}. The main difference is that here we do not require any predictive component related to $b$. For clarity, we present a simplified proof based on the similar approach from Theorem \ref{theorem:predictive safety}. From the proof of Proposition \ref{prop:predictive safety}, it is clear that for any $\myvar{x}_{k} \in \myset{C}(k)$, the local Lipschitz property of $h$ holds for $\myvar{x}_{k+1}$. Thus by the Lipschitz property, the following holds for any $k \in \mathbb{N}$: $|h(\myvar{f}(\myvar{x}_k, \myvar{u}_k,  k) + \myvar{d}(\myvar{x}_k, \myvar{u}_k, k), k+1) - h(\myvar{f}(\myvar{x}, \myvar{u}_k,k), k+1)| 
    \leq \Lip{h}{x}\| \myvar{f}(\myvar{x}_k, \myvar{u}_k, k) + \myvar{d}(\myvar{x}_k, \myvar{u}_k, k) - \myvar{f}(\myvar{x}, \myvar{u}_k, k) \| 
    = \Lip{h}{x} \| \myvar{d}(\myvar{x}_k, \myvar{u}_k, k)\| 
   \leq \Lip{h}{x} \Lip{d}{}$
It follows then that $h(\myvar{f}(\myvar{x}_k, \myvar{u}_k,k), k+1) - \Lip{h}{x} \Lip{d}{} \leq h(\myvar{f}(\myvar{x}_k, \myvar{u}_k,  k) + \myvar{d}(\myvar{x}_k, \myvar{u}_k, k), k+1) = h(\myvar{x}_{k+1}, k+1)$. Thus from \eqref{eq:dcbf predictive}, $h(\myvar{x}_{k+1}, k+1) \geq h(\myvar{f}(\myvar{x}_{k}, \myvar{u}_{k},k), k+1) - \Lip{h}{x} \Lip{d}{} \geq 0$. Thus $\myvar{x}_{k+1} \in \myset{C}(k+1)$. Now it is clear that since $\myvar{x}_{k_0} \in \myset{C}(k_0)$, then by induction $\myvar{x}_{k} \in \myset{C}(k) \subset \myset{X}(k)$ for all $k \geq k_0$.
\end{proof}

Next, we show that under stronger conditions, we can enforce safety by only implementing a control in $\myset{A}(k)$.
\begin{thm}\label{theorem:safety event trigger}
Suppose the conditions of Theorem \ref{theorem:safety} hold with $\myset{K}(\myvar{x}, k)$ defined by \eqref{eq:control safe set with A} in place of $\robset{K}$, $\myset{A}(k)$ defined by \eqref{eq:annulus set tv}, and $h$ is 1-step bounded on $\mathbb{N}$ with $\Lip{h}{k}\in \mathbb{R}_{\geq 0}$. Further suppose there exists $\Lip{f}{}\in \mathbb{R}_{>0}$ such that the following holds for all $\myvar{x} \in \myset{Y}$, $\myvar{u} \in \myset{U}$, $k \in \mathbb{N}$:
\begin{equation}\label{eq:f bound}
    \| \myvar{f}(\myvar{x}, \myvar{u}, k) - \myvar{x} \| \leq \Lip{f}{},
\end{equation}  
for which  $a \geq \Lip{h}{x} (\Lip{f}{}+\Lip{d}{}) + \Lip{h}{k}$, and the resulting $\myset{A}(k) \neq \emptyset $ $\forall k \in \mathbb{N}$. If $\myvar{x}_{k_0} \in \myset{C}(k_0) \subset \myset{X}(k_0)$ for some $k_0 \in \mathbb{N}$, and $\myvar{u}_k \in \myset{K}(\myvar{x}_k, k)$ is implemented in closed-loop with \eqref{eq:nonlinear system discrete}, then $\myvar{x}_{k} \in \myset{C}(k) \subset \myset{X}(k)$ for all $k\geq k_0, k \in \mathbb{N}$.
\end{thm}
\begin{proof}
Since the conditions of Theorem \ref{theorem:safety} hold on $\myset{A}(k) \subset \myset{C}(k)$, we need only consider the case when $\myvar{x}_k \in \myset{C}(k) \setminus \myset{A}(k)$ (i.e. $h(\myvar{x}_k, k) > a$). Let $\barvar{f}_k = \myvar{f}(\myvar{x}_k, \myvar{u}_k, k) + \myvar{d}(\myvar{x}_k, \myvar{u}_k, k)$ to simplify the notation, for which \eqref{eq:f bound} and the triangle inequality yields: $\|\barvar{f}_k - \myvar{x}_k\| = \| \myvar{f}(\myvar{x}_k, \myvar{u}_k, k) + \myvar{d}(\myvar{x}_k, \myvar{u}_k, k) - \myvar{x}_k\| \leq \|\myvar{f}(\myvar{x}_k, \myvar{u}_k, k)  - \myvar{x}_k\| + \|\myvar{d}(\myvar{x}_k, \myvar{u}_k, k) \| \leq \Lip{f}{} + \Lip{d}{}$. Using the following conditions: $|h(\myvar{y}, k) - h(\myvar{x}, k) | \leq \Lip{h}{x}\|\myvar{y} - \myvar{x}\|$, $|h(\myvar{x}, k+1) - h(\myvar{x}, k)| \leq \Lip{h}{k}$ and the triangle inequality, we arrive at the following inequality: $|h(\barvar{f}_k, k+1) - h(\myvar{x}_k, k )|  
    = |h(\barvar{f}_k, k+1) -h(\barvar{f}_k, k) + h(\barvar{f}_k, k) - h(\myvar{x}_k, k) | 
    \leq |h(\barvar{f}_k, k) - h(\barvar{x}_k, k)| + |h(\barvar{f}_k, k+1) - h(\barvar{f}_k, k)| 
     \leq \Lip{h}{x}\|\barvar{f}_k - \myvar{x}_k\| + \Lip{h}{k} 
    \leq \Lip{h}{x} (\Lip{f}{} + \Lip{d}{}) + \Lip{h}{k}  \leq a$
From this relation, it follows that $h(\myvar{x}_{k_0+1}, k_0+1) \geq h(\myvar{x}_{k_0}, k_0) - a > 0$. Since $\myvar{x}_{k_0} \in \myset{C}(k_0)$, then by induction $\myvar{x}_k \in \myset{C}(k)$ for all $k\geq k_0$.
\end{proof}

\begin{remark}
    The condition \eqref{eq:f bound} may seem highly restrictive at first, however many systems in fact satisfy this condition. For example, if $\myset{Y}$, i.e., the set that contains $\myset{X}(k)$ for all time, is compact and $\myvar{f}$ is time-invariant and a continuous function on its domain then in fact \eqref{eq:f bound} must hold, since $\myvar{x}$ and $\myvar{f}$ are both bounded on the respective sets. Continuity of $\myvar{f}$ is a common assumption for many discrete-time systems \cite{Wabersich2022a} and in many instances the safe set is compact \cite{Wabersich2022a, ShawCortez2022a}. In such cases, single integrators and mechanical/robotic systems \cite{ShawCortez2022a} satisfy \eqref{eq:f bound}. The 1-step bound condition only restricts how much $h$ can change with respect to time, between time steps. This condition is needed to not require checking the barrier condition in $\myset{C}(k)\setminus \myset{A}(k)$.
\end{remark}
\begin{remark}
Theorem \ref{theorem:safety event trigger} provides a new condition of forward invariance for discrete-time systems that does not require checking the condition \eqref{eq:dcbf predictive} for all states in $\myset{C}(k)$. This extends the standard results of forward invariance from \cite{Blanchini2015} (for which Theorem \ref{theorem:safety event trigger} provides a sufficient condition) by requiring stronger conditions of $h$ and the system dynamics. Theorem \ref{theorem:safety event trigger} provides a similar notion of forward invariance used in continuous-time systems, for which one need-only check how the system behaves near the constraint boundary to ensure safety.
\end{remark}

The predictive safety control can be implemented in a 1-step look-ahead fashion, typical of many existing barrier function methods. In this case, the proposed control can be simplified into the following optimization problem:
 \begin{subequations}\label{eq:psf original 1step}
\begin{flalign}
    & \myset{P}_{sf}(\myvar{x}_k, k)= \nonumber &&\\
  \vspace{0.1cm}  &  \hspace{0.5cm}  \underset{\myvar{u}_k } {\text{argmin}} \hspace{.3cm} \|\myvar{u}_{nom}(\myvar{x}_k, k) - \myvar{u}_k\|  &&\\
  & \hspace{0.5cm} \text{s.t. }   &&\\
& \hspace{0.5cm}  \robvar{x}_{0|k} = \myvar{x}_k,  &&\\
& \hspace{0.5cm}  \robvar{x}_{1|k} = \myvar{f}(\robvar{x}_{0|k}, \myvar{u}_k,k),   &&\\
& \hspace{0.5cm}  \myvar{u}_k \in \myset{U}, &&\\
& \hspace{0.5cm}  h(\robvar{x}_{1|k}, k+1) \geq \Lip{h}{x}\Lip{d}{}  \label{eq:psf original terminal 1step} &&
\end{flalign}
\end{subequations}

In the special case where the conditions of Theorem \ref{theorem:safety event trigger} hold, the simpler Algorithm \ref{alg:event-triggered control 1 step} can be implemented. Note that Algorithm \ref{alg:event-triggered control 1 step} is not the same as Algorithm \ref{alg:event-triggered control} with $N=1$. In Algorithm \ref{alg:event-triggered control 1 step}, no system dynamics roll-out is required. The only condition that needs to be checked is if $h(\myvar{x}_k,k) > a$.

\begin{algorithm}[!htbp]
\caption{1-Step Event-Triggered, Robust Safety Filter}\label{alg:event-triggered control 1 step}
\begin{algorithmic}[1]
\State Given: $\myvar{x}_k$, $k$.
\If{ $h(\myvar{x}_k, k) > a$}
    \State \Return $\myvar{u}_k = \myvar{u}_{nom}(\myvar{x}_k, k)$.
\Else
    \State Compute \eqref{eq:safety filter control} with  $\myset{P}_{sf}$ from \eqref{eq:psf original 1step} for $\myvar{u}_k^*$.
    \State \Return $\myvar{u}_k = \myvar{u}_k^*$.
\EndIf
\end{algorithmic}
\end{algorithm}

\begin{cor}\label{cor:safety event trigger 1step}
Suppose the conditions of Theorem \ref{theorem:safety event trigger} hold. Given a nominal control law $\myvar{u}_{nom}: \mathbb{R}^{n} \times \mathbb{N} \to \myset{U}$, suppose the system \eqref{eq:nonlinear system discrete} is in closed-loop with the control from Algorithm \ref{alg:event-triggered control 1 step}. Then if $\myvar{x}_{k_0} \in \myset{C}(k_0) \subset \myset{X}(k_0)$ for any $k_0\in \mathbb{N}$, then $\myvar{x}_{k} \in \myset{C}(k) \subset \myset{X}(k),\ \forall k\geq k_0, k \in \mathbb{N}$.
\end{cor}
\begin{proof}
Follows from Theorem \ref{theorem:safety event trigger}.
\end{proof}

\section{Numerical Examples}\label{sec:examples}

In this section, the proposed controllers from Algorithm \ref{alg:event-triggered control} and \ref{alg:event-triggered control 1 step} are implemented on several systems. 
The nominal control used was the learning-based differentiable predictive control (DPC) from \cite{Drgona2024,Drgona2022c}. The DPC method in all presented examples was designed and trained using the Neuromancer scientific machine learning library~\cite{Neuromancer2023}. All code was developed in Python using the CasADi optimization module~\cite{Andersson2019} to solve $\myset{P}_{sf}$. For each example, the DPC control policy was parametrized by a multi-layer perceptron (MLP) of 2 layers with 32 internal states each and a Gaussian Error Linear Unit (GELU) activation. The MLP was implemented with a sigmoid scale method to ensure input constraints were satisfied.

\subsection{Two-Tank Example}\label{ssec:two tank example}

We consider the nonlinear, discretized two-tank system
: $\bar{f}_1(\myvar{x}, \myvar{u}) := x_{1} + \Delta t\left( c_1(1-u_2)u_1 - c_2\sqrt{x_{1}} \right) $, $\bar{f}_2(\myvar{x}, \myvar{u}) := x_{2} + \Delta t \left( c_1u_1u_2 + c_2\sqrt{x_{1}} - c_2\sqrt{x_{2}} \right)  $
where $x_1, x_2 \in \mathbb{R}_{>0}$ are the height of the liquid in tank 1 and 2, respectively, $c_1 = 0.8$ is the inlet valve coefficient, $c_2 = 0.4$ is the outlet valve coefficient, and $u_1, u_2 \in \mathbb{R}_{\geq 0}$ are the pump and valve control terms, respectively. The sampling time used was $\Delta t = 0.1$ s. The objective is to keep the liquid between desired levels in each tank i.e., $\myset{X} = \{ \myvar{x} \in \mathbb{R}^2: 0.2\leq x_i\leq 1, \ i\in \{1,2\} \}$ for $\myvar{x} = [x_1, x_2]^T$, while respecting the input constraints defined by: $\myset{U}=\{\myvar{u}: 0\leq u_i \leq 1.0, \forall i\in \{1,2\}\}$ and tracking a piece-wise constant reference trajectory $\myvar{r}: \mathbb{N}\to \mathbb{R}^n$. 

Note that some optimization modules, such as CasADi, have difficulty when non-smooth dynamics are included in the optimization. However, thanks to the robust nature of the proposed control, we re-write the two tank system with an approximation of $\sqrt{\cdot}$ as: $f_1(\myvar{x}, \myvar{u}) = x_1 + \Delta t\left(c_1(1-u_2)u_1 - c_2 \psi(x_1) \right),$ $
    f_2(\myvar{x}, \myvar{u}) = x_2 + \Delta t \left(c_1u_1u_2 + c_2\psi(x_1) - c_2\psi(x_2)  \right)$, 
where $\psi:\mathbb{R}\to\mathbb{R}$ is a 7th order polynomial fit of $\sqrt{x}$ on $[0.2,1]$, which yields:
\begin{align*}
    x_{1_{k+1}} &= f_1(\myvar{x}_k, \myvar{u}_k) + (\bar{f}_1(\myvar{x}_k, \myvar{u}_k) - f_1(\myvar{x}_k, \myvar{u}_k)) + w_1(k) \\
    x_{2_{k+1}} &= \underbrace{f_2(\myvar{x}_k, \myvar{u}_k)}_{\myvar{f}(\myvar{x}_k, \myvar{u}_k)} + \underbrace{(\bar{f}_2(\myvar{x}_k, \myvar{u}_k) - f_2(\myvar{x}_k, \myvar{u}_k)) + w_2(k)}_{\myvar{d}(\myvar{x}_k, \myvar{u}_k, k)}
\end{align*}
where $w_1(k) = w_2(k) = \bar{w} sin(k)$ are additional perturbations acting on the system for $\bar{w} \in \mathbb{R}_{>0}$.
We define $h(\myvar{x}) = \varepsilon - (x_1 - x_{r_1})^2 + \rho(x_2 - x_{r_2})^2$, $\myvar{x}_r =[x_{r_1} = 0.63, x_{r_2} = 0.63]^T$, $b_1(\myvar{x}) = x_{max} - x_1$, $b_2(\myvar{x}) = x_1 - x_{min}$, $b_3(\myvar{x}) = x_{max} - x_2$,  $b_4(\myvar{x}) = x_2 - x_{min}$, where $x_{min} = 0.2$, $x_{max} = 1.0$, $\rho = 2.69$, and $b_1, b_2, b_3, b_4$ are used as discussed in Remark \ref{rem:multiple constraints} and note that these all have the same Lipschitz constant $\Lip{b}{x}$. The function $h$ is shown to be an \emph{$N$-step} \RBF \hspace{0.1mm} in Appendix \ref{app:two tank barrier check} for $N=20$. The parameters used for the first simulation of the two-tank are as follows: $\bar{w} = 0.00001$, $\Lip{f}{x} = 1.205$, $\Lip{b}{x} = 1.0$, $\Lip{h}{x} = 1.331$, $\Lip{d}{} = 0.0000542$, $\varepsilon = 0.12$ and $\myvar{x}_r = [0.63, 0.63]^T$. Note that $\Lip{d}{}$ was determined by combining $\bar{w}$, the additional perturbation, with the computed bounded error between $\psi(x)$ and $\sqrt{x}$ on $[0.2, 1.0]$. The DPC control law for the two-tank system was designed  to track the time-varying reference and penalty functions were used to address the system constraints. 

\begin{figure}
    \centering
    \begin{subfigure}{0.5\textwidth}
        \centering
        \includegraphics[width=\textwidth]{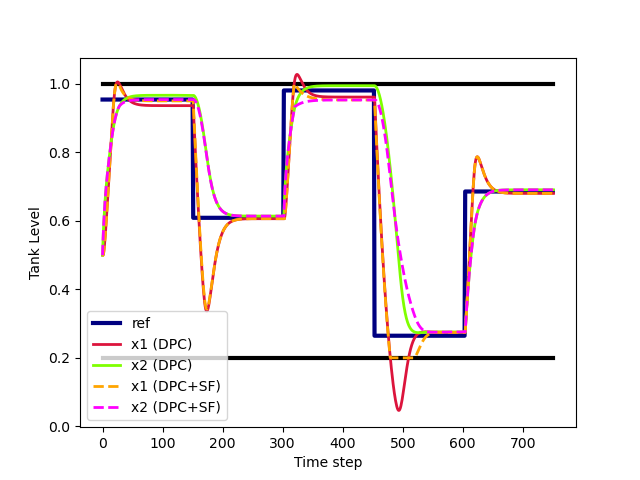}
        \caption{State trajectories with constraint bounds (black, solid lines).}
        \label{fig:two_tank_state_traj}
    \end{subfigure}

     \begin{subfigure}{0.5\textwidth}
        \centering
        \includegraphics[width=\textwidth]{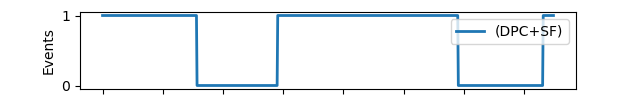}
        \caption{Events triggered (1 is triggered, 0 is not triggered) for each simulation scenario. Note the `DPC' scenario is not shown as no safety filter is implemented.}
        \label{fig:two_tank_events}
    \end{subfigure}

     \begin{subfigure}{0.5\textwidth}
        \centering
        \includegraphics[width=\textwidth]{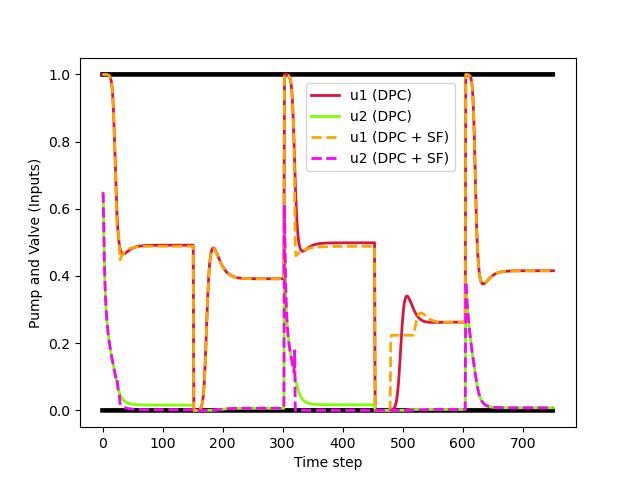}
        \caption{Input trajectories with constraint bounds (black, solid lines).}
        \label{fig:two_tank_input_traj}
    \end{subfigure}
    
    \caption{(Two-tank example) Comparison of trajectories for `DPC' and `DPC+SF' implementations with the proposed robust safety filter on the perturbed two-tank system with $\bar{w} = 0.00001$.}
    \label{fig:two_sim_results1}
\end{figure}

In the first set of simulations, the DPC control, denoted `DPC' was implemented alone in one scenario and the proposed control \eqref{alg:event-triggered control}, denoted `DPC+SF', was implemented in the second scenario with $N = 20$. For the chosen $N$, the conditions of Assumption \ref{asm:nonempty sets} hold. The results of the simulation are shown in Figure \ref{fig:two_sim_results1}.

The plots in Figure \ref{fig:two_sim_results1} show that the `DPC' control is able to track the reference trajectory reasonably well, but violates the system constraints. 
The constraint violations occur at time steps $k = 24$, $324$ and $493$. The `DPC+SF' control, on the other hand, implements the `DPC' control as close as possible, but then overrides the `DPC' control (see Figure \ref{fig:two_tank_input_traj}) to ensure the states remain inside their constraint bounds (see Figure \ref{fig:two_tank_state_traj}). Furthermore, the event-triggering from Algorithm \ref{alg:event-triggered control} is shown in Figure \ref{fig:two_tank_events} for which the optimization problem from the safety filter is only solved when the system states move close to the constraint boundary. These results show the advantage of the proposed methodology from Algorithm \ref{alg:event-triggered control} in enforcing system constraints without requiring the optimization problems to be solved at every instant when implemented online.

\begin{figure}
    \centering
    \begin{subfigure}{0.5\textwidth}
        \centering
        \includegraphics[width=\textwidth]{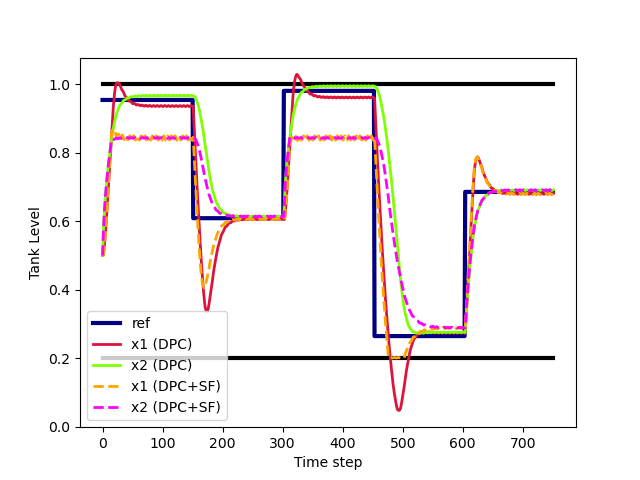}
        \caption{State trajectories with constraint bounds (black, solid lines).}
        \label{fig:two_tank_state_traj2}
    \end{subfigure}

     \begin{subfigure}{0.5\textwidth}
        \centering
        \includegraphics[width=\textwidth]{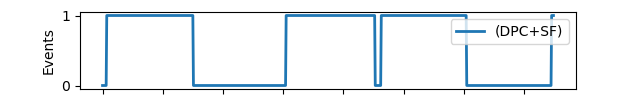}
        \caption{Events triggered (1 is triggered, 0 is not triggered) for each simulation scenario. Note the `DPC' scenario is not shown as no safety filter is implemented.}
        \label{fig:two_tank_events2}
    \end{subfigure}

     \begin{subfigure}{0.5\textwidth}
        \centering
        \includegraphics[width=\textwidth]{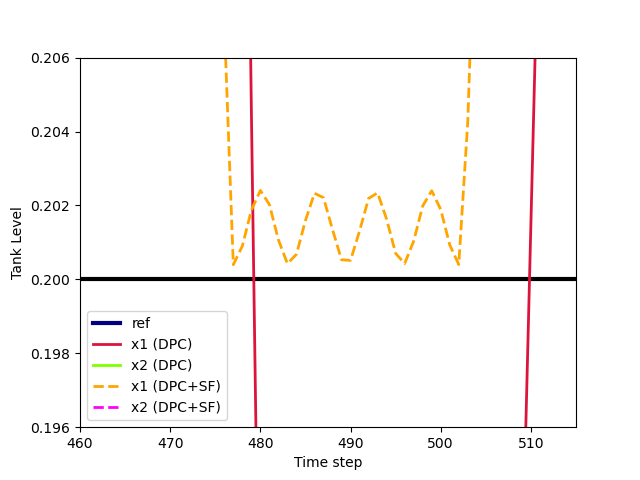}
        \caption{State trajectories (zoomed-in) with constraint bounds (black, solid lines). Shows proposed safety filter satisfying system constraints.}
        \label{fig:two_tank_state_traj2_zoomed}
    \end{subfigure}

    
    \caption{(Two-tank example) Comparison of trajectories for `DPC' and `DPC+SF' implementations with the proposed robust safety filter on the perturbed two-tank system with $\bar{w} = 0.001$.}
    \label{fig:two_sim_results2}
\end{figure}

Although these results are promising, there is a trade-off between the size of the optimization problems from the safety filter and the magnitude of the disturbance that can be tolerated by the system. In other words, as $N$ increases the robustness margins from \eqref{eq:robust safe set} and \eqref{eq:robust safe terminal set} increase, which restricts the magnitude of $\Lip{d}{}$ that can be tolerated, and vice versa. To show this, we implemented the proposed control with a higher perturbation, $\bar{w} = 0.001$ for which $\Lip{d}{} = 0.00145$, and resulted in a smaller prediction horizon of $N=6$. The results of this simulation are shown in Figure \ref{fig:two_sim_results2}. The plots show more conservative behavior in the `DPC+SF' case compared to the previous simulation, especially between $k=14$ and $k=150$ and $k=301$ and $k=451$. Within both of these time intervals, the shorter prediction horizon requires the system to stay closer to $\myset{C}$ to satisfy the safety conditions. Note however that the proposed control is able to reject the larger perturbation here and keep the system safe. 

\begin{figure}
    \centering
    \begin{subfigure}{0.5\textwidth}
        \centering
        \includegraphics[width=\textwidth]{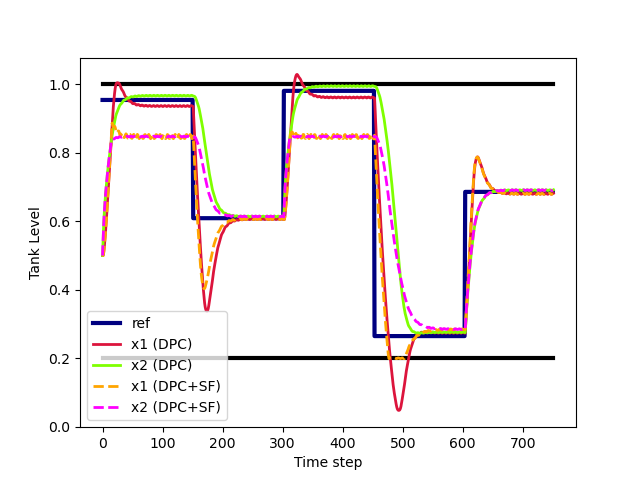}
        \caption{State trajectories with constraint bounds (black, solid lines).}
        \label{fig:two_tank_state_traj3}
    \end{subfigure}

     \begin{subfigure}{0.5\textwidth}
        \centering
        \includegraphics[width=\textwidth]{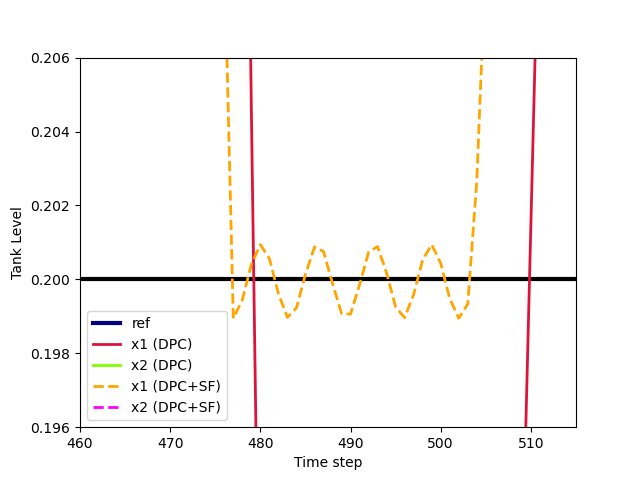}
        \caption{State trajectories (zoomed-in) with constraint bounds (black, solid lines). Original safety filter violates safety constraints.}
        \label{fig:two_tank_state_traj3_zoomed}
    \end{subfigure}

    
    \caption{(Two-tank example) Results for predictive safety filter of \cite{Wabersich2022a} for `DPC' and `DPC+SF' implementations on the perturbed two-tank system with $\bar{w} = 0.001$.}
    \label{fig:two_sim_results3}
\end{figure}

Finally, to showcase the approach compared to the existing predictive safety filter of \cite{Wabersich2022a}, we implement that predictive safety filter on the previous example with $\bar{w} = 0.001$. 
The same parameters used for the safety filter from Algorithm \ref{alg:event-triggered control} were used for the safety filter of \cite{Wabersich2022a} for a direct comparison in addition to $\Delta = 0.00000005$ (see \cite{Wabersich2022a}). The results of the safety filter implementation are shown in Figure \ref{fig:two_sim_results3}, which shows similar performance to the proposed control in Figure \ref{fig:two_sim_results2}, except for the state constraint violation that occurs between $k=476$ and $k=53$ (see Figure \ref{fig:two_tank_state_traj3_zoomed}). Furthermore, the control from \cite{Wabersich2022a} does not include an event-triggering and so must be solved at all instances in time. As a result, the proposed control was 66.0\% faster in online implementation compared to that of \cite{Wabersich2022a} (see Figure \ref{fig:two_tank_events2}). These results show that the proposed control is beneficial for handling bounded perturbations to guarantee hard safety constraints at all times, while reducing the online computations required.

\subsection{Building Example}

\begin{figure}
    \centering
    \includegraphics[width=0.5\textwidth]{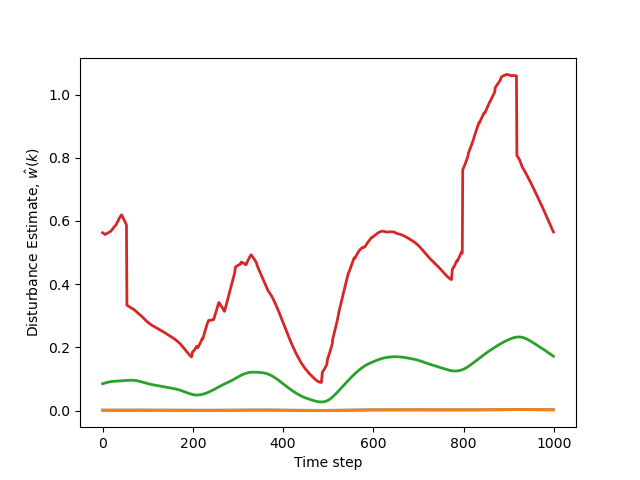}
    \caption{Estimate of environmental perturbations on the building, $\robvar{w}(k)$.}
    \label{fig:building disturbance}
\end{figure}

Here we apply the proposed method to a single zone building temperature control problem from~\cite{Drgona2013}. This is a nonlinear, time-varying system that incorporates an estimate of the ambient disturbance affecting the building, defined as follows: $\myvar{x}_{k+1} = \underbrace{A\myvar{x}_k + Bu_k + \robvar{w}(k)}_{\myvar{f}(\myvar{x}_k, \myvar{u}_k, k)} + \underbrace{(\myvar{w}(k) - \robvar{w}(k))}_{\myvar{d}(k)} $, with
\begin{equation*}
    A = \left[\begin{matrix}
        0.995 &  0.0017 &0.0 &     0.0031 \\
 0.0007 & 0.996 & 0.0003 & 0.0031 \\
 0.0 &     0.0003 & 0.983 & 0.0    \\
 0.202 & 0.488& 0.01  & 0.257
    \end{matrix} \right], 
    B = \left[\begin{matrix}
        0.00000176 \\
 0.00000176 \\
 0.0        \\
 0.000506
    \end{matrix} \right]
\end{equation*}
where $\myvar{w}:\mathbb{N}\to \mathbb{R}^n$ is the perturbation of the environment affecting the system, and $\robvar{w}:\mathbb{N}\to \mathbb{R}^n$ is the estimate of $\myvar{w}$, which is commonly approximated using annual statistical information of the building and surrounding region (see Figure \ref{fig:building disturbance}).  The control $u$ is the heat flow of the HVAC system with input constraints set: $\myset{U} = \{u\in\mathbb{R}: 0\leq u\leq 5000 \}$, and the number of system states is $n=6$. Note that these dynamics were developed for a sampling rate of 0.1 s. For this example, $\Lip{d}{} = 0.005$ and $\myvar{d}(k) = -\Lip{d}{} \myvar{1}_n $ was used to demonstrate the worst-case disturbance that is constantly lowering the temperature of the building.

The objective is to minimize the energy usage, i.e., the control input, while remaining within time-varying constraints defined by: $b_1(\myvar{x},k) = -C \myvar{x}  + \bar{r}(k)$, $b_2(\myvar{x},k) = C \myvar{x}  - \ubar{r}(k)$, where $C = [0, 0, 0, 1]$ and $\bar{r}, \ubar{r}: \mathbb{N} \to \mathbb{R}$ define the time-varying comfort constraint bounds. The constraint functions were implemented as per Remark \ref{rem:multiple constraints}. 

Here we take a different approach to the previous example and show how the proposed control could be implemented from a practical perspective. First, it is reasonable to assume that the building HVAC system is capable of keeping the building room temperature within a temperature range despite reasonable fluctuations in the ambient temperature. Here we use a temperature range from 19.55 C to 20.45 C. We also assume that if the output of the system, $C \myvar{x}$, which represents the temperature of the simple single zone, is bounded, then the states are also bounded. Thus we effectively assume the following function is a \emph{N-step} \RBF \hspace{0.1mm}: $h(\myvar{x}) = \varepsilon - \|C \myvar{x}  - y_r\|_2^2$, where $\varepsilon = 0.9$, $y_r = 20$ C is the reference temperature, and $N=6$. Note that neither the state constraint sets: $\myset{X}(k)$
nor the terminal constraint set $\myset{C}$ are compact, however compactness is not required for the results of Theorem \ref{theorem:predictive safety}. We define the robustness margins used: $\Lip{f}{x} = 0.9998$, $\Lip{b}{x} = 1.0$, $\Lip{h}{x} = 100$. The margins for $\Lip{f}{x}$ and $\Lip{b}{x}$ were computed using standard methods of computing Lipschitz constants \cite{Khalil2002}. For $\Lip{h}{x}$, since $h$ is quadratic, we use the assumption that the states remain bounded if the temperature is bounded in the terminal set to make a conservative estimate of the bound of $\|\myvar{x}\|$ for all $\myvar{x} \in \myset{C}$. 

\begin{figure}
    \centering
    \includegraphics[width=0.5\textwidth]{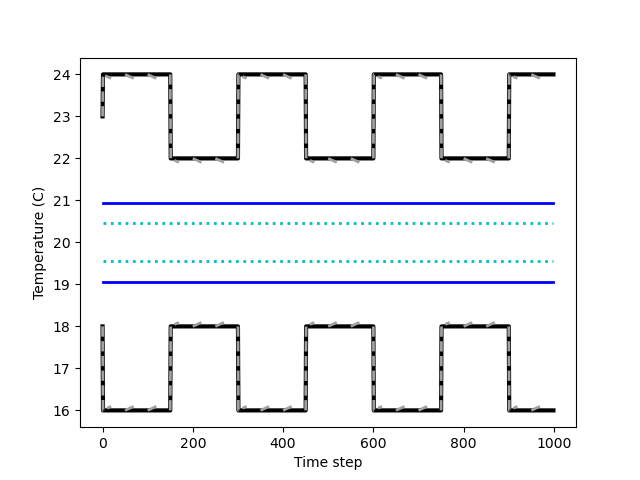}
        \caption{(Building example) Comfort constraints (black, solid lines) along with terminal constraint (blue, solid lines). The sets from Assumption \ref{asm:nonempty sets} are depicted as: $\myset{X}^l(k+l)$ for $l\in \{0,...,N\}$ is the region between the gray, dashed lines depicted at intervals of 50 time steps, $\myset{X}_{f}^N(k)$ is the region between the cyan, dotted lines. 
        }
        \label{fig:buildling sets}
\end{figure}

From Figure \ref{fig:buildling sets}, it is clear that the sets from Assumption \ref{asm:nonempty sets} are non-empty and that the robust safety set with respect to the terminal barrier function, i.e., $\myset{X}_{f}^N(k+N)$, is a subset of the robust safety set with respect to the state constraints, i.e., $\myset{X}^N(k+N)$. Thus the conditions of Assumption \ref{asm:nonempty sets} are satisfied. The DPC control law was designed using a regulation loss term to minimize the control input of the system. In addition, a smoothness term was added to prevent sharp changes in the control input, and a penalty loss was used to address state constraints (see \cite{Drgona2024}).

\begin{figure}
    \centering
    \begin{subfigure}{0.5\textwidth}
        \centering
        \includegraphics[width=\textwidth]{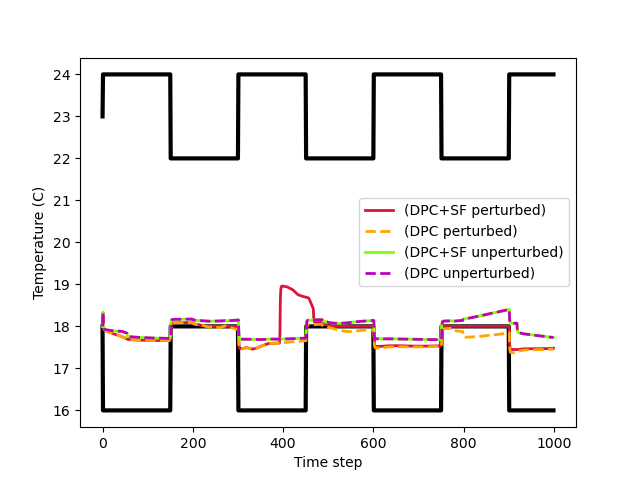}
        \caption{Temperature trajectories with comfort constraints (black, solid lines).}
        \label{fig:building_state_traj}
    \end{subfigure}
    
    \begin{subfigure}{0.5\textwidth}
        \centering
        \includegraphics[width=\textwidth]{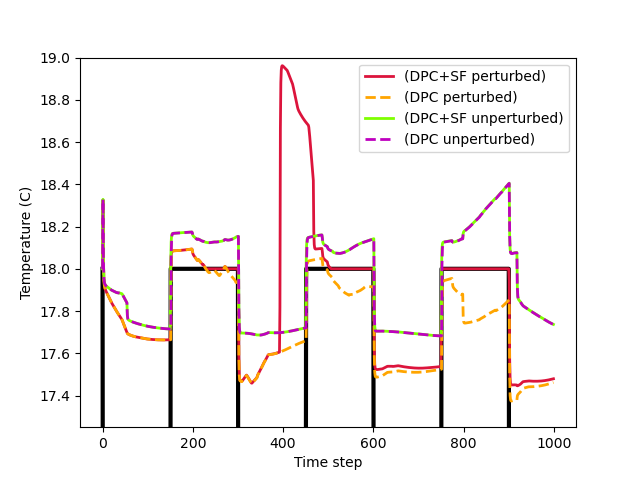}
        \caption{Temperature trajectories (zoomed-in) with comfort constraints (black, solid lines).}
        \label{fig:building_state_traj_zoomed}
    \end{subfigure}

     \begin{subfigure}{0.5\textwidth}
        \centering
        \includegraphics[width=\textwidth]{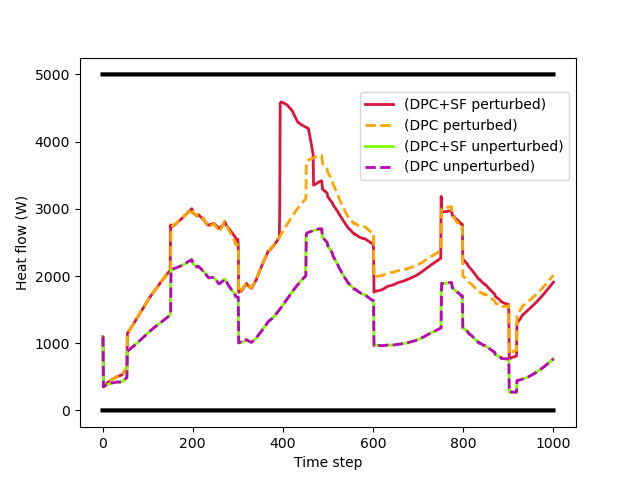}
        \caption{Input trajectories with constraint bounds (black, solid lines).}
        \label{fig:building_input_traj}
    \end{subfigure}
    
    \caption{(Building example) Comparison of trajectories for `DPC+SF' with perturbation, `DPC' with perturbation, `DPC+SF' without perturbation, and `DPC' without pertubation.}
    \label{fig:buildling_sim_results}
\end{figure}

The proposed control from Algorithm \ref{alg:event-triggered control} is compared to using the DPC control alone. We refer to the control from Algorithm \ref{alg:event-triggered control} as `DPC+SF', and DPC as simply `DPC'. Both controllers were run with and without the perturbation term $\myvar{d}(k)$. The results are shown in Figure \ref{fig:buildling_sim_results}. 

Figure \ref{fig:building_state_traj} shows the comparisons of the various different simulations. The unperturbed cases show that the `DPC+SF' and `DPC' controllers overlap throughout the entire simulation in both the temperature and input trajectories. This indicates that in the absence of uncertain disturbances, DPC is able to respect the safety constraints. If the safety constraints were ever to be violated, the safety filter would have altered the DPC control. This result is expected as DPC is known for (probabilistically) ensuring constraint satisfaction in the abscence of perturbations. This is not the case with the perturbed simulations. The trajectories associated with the dynamics subject to $\myvar{d}(k)$ show that the safety filter (see the `DPC+SF' trajectories) must alter the original DPC control to remain within the comfort constraints. The `DPC+SF' remains within both the comfort constraints and input constraints. The `DPC' control on the other hand violates the comfort constraints (see Figure \ref{fig:building_state_traj}). These results show that the proposed control, `DPC+SF' is able to robustly handle uncertain perturbations, while remaining point-wise, minimally close to the DPC control law. Note that in this example, the objective was to reduce the amount of energy used, while staying within the comfort constraints, which resulted in trajectories close to the system constraints. As a result, the simulation resulted in events triggering at every time step. This highlights an important aspect of the event-triggering component in that if the nominal control aims to keep the system close to the constraint bounds, the safety filter will need to be solved at all time steps.

\subsection{Simple Example (1-step Safety Filter)}

In this simple example, we apply Algorithm \ref{alg:event-triggered control 1 step} to the perturbed single integrator discretized using the standard Euler method: $x_{k+1} =\underbrace{ x_k + \Delta t(u_k) }_{f(x_k, u_k,k)} +\underbrace{\Delta t w(k)}_{d(x_k, u_k, k)}$,  
where $\Delta t =0.01$ s is the sampling time and $w(k) = 0.02 sin(0.5 k)$ is a perturbation on the system. 
The input constraint set is: $\myset{U} = \{u\in \mathbb{R}: |u| \leq 10\}$. The barrier function used here is: $ h(x, k) = \varepsilon - (x - x_r(k))^2$, for $x_r(k)=0.5 sin( 0.05 k)$
The objective is to keep the system inside the time-varying tube defined by $\myset{C}(k)\subseteq \myset{X}(k)$, similar to the continuous-time version from \cite{ShawCortez2022a}.

\begin{figure}
    \centering
    \includegraphics[width=0.5\textwidth]{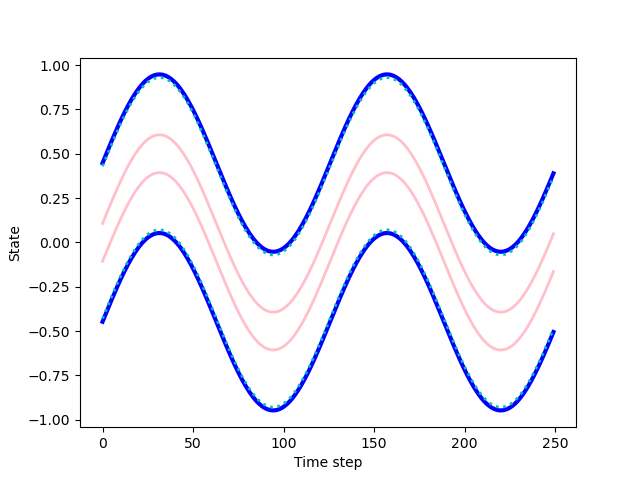}
        \caption{(Simple example) The barrier function safe set, $\myset{C}(k)$, is the region between the blue, solid lines. The set $\myset{X}_{f}^N(k)$ is the region between the cyan, dotted lines. The set $\myset{A}(k)$ is the region between the pink solid lines and the blue solid lines.}
        \label{fig:single integrator sets}
\end{figure}

To apply Algorithm \ref{alg:event-triggered control 1 step}, we need to ensure that $h$ is a \emph{1-step} \RBF \hspace{0.1mm} and that the conditions of Theorem \ref{theorem:safety event trigger} hold. The Lipschitz constants and bounds from Theorem \ref{theorem:safety event trigger} are computed by using standard methods for which $\Lip{h}{x} = 0.894$, $\Lip{h}{k} = 0.072$, $\Lip{d}{}=0.02$, $\Lip{f}{} = 0.11$, $\Lip{f}{x} = 1.0$. See Appendix \ref{app:simple integrator barrier check} to check that $h$ is a \emph{1-step} \RBF. We chose $a = \Lip{h}{x}(\Lip{f}{} + \Lip{d}{}) + \Lip{h}{k} = 0.180$. Since $0<a < \varepsilon$ and $a > 0$, $\myset{A}(k) \neq \emptyset$ and $\myset{A}(k) \subset \myset{C}(k)$. All the conditions of Theorem \ref{theorem:safety event trigger} have been met. Figure \ref{fig:single integrator sets} shows the sets associated with this example.
In this example, the DPC control was used to reduce the amount of online computations needed to satisfy the objective. DPC was tasked with tracking the center of the tube to avoid triggering the event in Algorithm \ref{alg:event-triggered control 1 step}. 

\begin{figure}
    \centering
    \begin{subfigure}{0.5\textwidth}
        \centering
        \includegraphics[width=\textwidth]{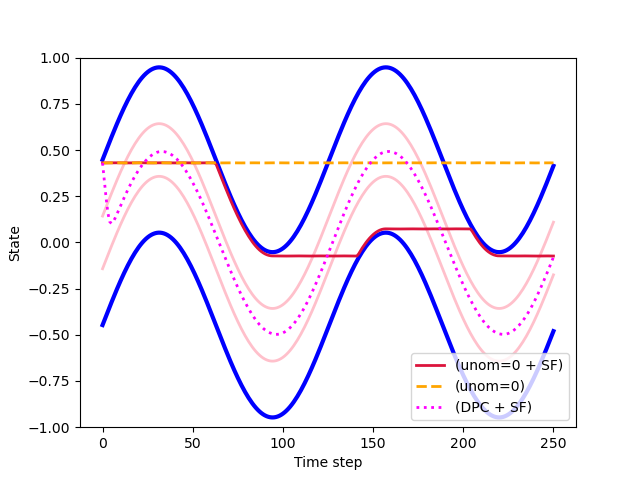}
        \caption{State trajectories with barrier function safe set (blue, solid lines).}
        \label{fig:single_integrator_state_traj}
    \end{subfigure}

     \begin{subfigure}{0.5\textwidth}
        \centering
        \includegraphics[width=\textwidth]{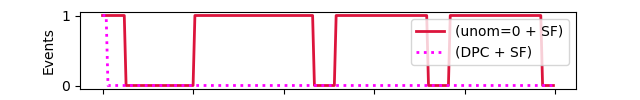}
        \caption{Events triggered (1 is triggered, 0 is not triggered) for each simulation scenario. Note the case where $u_{nom} = 0$ is not shown as no safety filter is implemented.}
        \label{fig:single_integrator_events}
    \end{subfigure}

     \begin{subfigure}{0.5\textwidth}
        \centering
        \includegraphics[width=\textwidth]{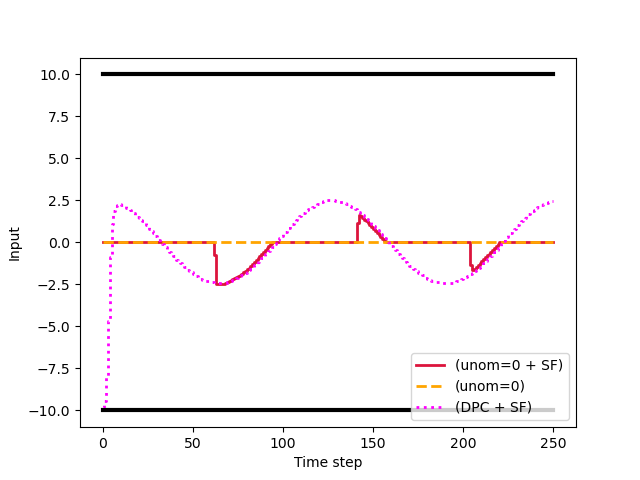}
        \caption{Input trajectories with constraint bounds (black, solid lines).}
        \label{fig:single_integrator_input_traj}
    \end{subfigure}
    
    \caption{(Simple example) Comparison of trajectories for `unom=0 + SF' (red, solid curve), `unom=0' (orange, dashed curve), `DPC+SF' (magenta, dotted curve).}
    \label{fig:single_integrator_sim_results}
\end{figure}

Algorithm \ref{alg:event-triggered control 1 step} was implemented for this example along with a `no control' case denoted `unom = 0'. This yields three different controllers. First is the `unom = 0' control in which $u(x,k)=u_{nom}(x,k)=0$ is implemented in closed-loop with the system. Second is the `unom = 0 + SF' in which Algorithm \ref{alg:event-triggered control 1 step} is implemented with $u_{nom}(x,k) = 0$. Third is the `DPC + SF' in which Algorithm \ref{alg:event-triggered control 1 step} is implemented as stated. Figure \ref{fig:single_integrator_sim_results} shows the results of these three scenarios.

In Figure \ref{fig:single_integrator_sim_results}, several concepts are demonstrated. First, it is clear that the `unom = 0' case is not able to stay within the tube as it quickly leaves the safe set. The same control implemented in Algorithm \ref{alg:event-triggered control 1 step}, i.e., `unom = 0 + SF', matches the `unom = 0' control until the system attempts to leave the safe region and the safety filter acts to keep the system within the safe set. This demonstrates the basic concept that the safety filter can ensure constraint satisfaction.

Second, the results show the impact of combining the DPC control with the SF in Algorithm \ref{alg:event-triggered control 1 step}. In the `unom = 0 + SF' case, the safety filter gets triggered frequently which requires the system to continually solve the nonlinear program online. This is shown in Figure \ref{fig:single_integrator_events}. Note that the events only occur when the state enters $\myset{A}(k)$. In the `DPC + SF' case, the safety filter is only solved within the first few time steps as the initial condition lies inside of $\myset{A}(k)$. However the DPC control is able to keep the system centered in the safe set which results in no further events triggering for the duration of the simulation. This shows that combining DPC, i.e., a learning-based control, with the safety filter allows for significant reduction in online computations, while still providing guarantees of safety in the presence of perturbations.

\section{Conclusion}

In this paper, we present a robust predictive safety filter to provide a robustly safe control law. The approach extends the existing state-of-the-art by handling unknown perturbations and extending the safety results to time-varying systems with time-varying constraints. Furthermore, the proposed method uses event-triggering to reduce the amount of online computation required. We discuss trade-offs in the design of the proposed control and implement it on a simple single integrator, a two-tank system, and a building system to demonstrate its efficacy.

The approach is dependent on the existence of an \emph{N-step} \RBF, which for general time-varying systems is not straightforward to design. Existing methods can be used to design barrier functions for the time-invariant case, but a general method for time-varying systems is still an open problem and left for future work. Additionally, future work will extend the methodology to include adaptive model updates for reduced conservatism.

\section{Appendix}

\subsection{Barrier function check for the simple integrator }\label{app:simple integrator barrier check}

First, we define the following control to satisfy \eqref{eq:dcbf predictive}:
\begin{equation}\label{eq:single integrator control}
    \hat{u}(x,k) = \begin{cases}
    0, \text{ if }  h(f(x,0,k),k+1) \geq \Lip{h}{x}\Lip{d}{},\\
        y(x,k) - \text{sign}(y(x,k)) \sqrt{\varepsilon - \Lip{h}{x} \Lip{d}{}}, \text{ otherwise} 
    \end{cases}
\end{equation}
where $y(x,k) = \frac{1}{\Delta t}(-x + x_r(k+1))$. We leave it to the reader to see that \eqref{eq:single integrator control} satisfies \eqref{eq:dcbf predictive} for all $x \in \myset{C}(k)$, $\forall k \in \mathbb{N}$. Next we need to make sure this control satisfies the input constraints for which it can be shown that $|\hat{u}| \leq \frac{1}{\Delta t}(\Lip{h}{k} + \sqrt{\varepsilon}- \sqrt{\varepsilon-\Lip{h}{x}{}\bar{w}}) \leq 9.28$. Thus $\hat{u} \in \myset{U}$. Thus we have shown that $h$ is a \emph{1-step} \RBF. 

\subsection{Barrier function check for the two-tank system} \label{app:two tank barrier check}

To show that $h$ is a barrier function, we use a sampling-based approach.
We first, define a feasible control to attempt to satisfy \eqref{eq:dcbf predictive}. This control consists of the following components: $y_1(\myvar{x}) =  -\frac{1}{\Delta t}(x_1 - x_{r_1})  + c_2 \psi(x_1), 
    y_2(\myvar{x}) =  -\frac{1}{\Delta t}(x_2 - x_{r_2})  - c_2 \psi(x_1) + c_2\psi(x_2) $.
 We can now define the proposed control noting that the condition \eqref{eq:dcbf predictive} is only required when $\myvar{x}_k \in \myset{C}$. The condition \eqref{eq:dcbf predictive} can be written as: $h(\myvar{f}(\myvar{x}, \myvar{u})) = \varepsilon - \Delta t^2(c_1(1-u_2)u_1 - y_1(\myvar{x}))^2 - \rho \Delta t^2(c_1u_1u_2 - y_2(\myvar{x}))^2 \geq \Lip{h}{x} \Lip{d}{} (\Lip{f}{x})^{N-1}$. Thus we can define an optimal point-wise control to satisfy the condition by: $ \robvar{u}(\myvar{x}) = \min_{\myvar{u} \in \myset{U}} \left(c_1(1-u_2)u_2 - y_1(\myvar{x})\right)^2 + \rho\left( c_1u_1u_2 - y_2(\myvar{x})\right)^2 $.
 The chosen control law to show that $h$ is a barrier function is:
 \begin{equation*}
     \barvar{u}(\myvar{x}) = \begin{cases}
         0, \text{ if } h(\myvar{f}(\myvar{x}_{k}, 0)) \geq \Lip{h}{x} \Lip{d}{} (\Lip{f}{x})^{N-1}, \\
         \robvar{u}(\myvar{x}), \text{ otherwise }
     \end{cases}
 \end{equation*}
 The control $\barvar{u}$ was checked over a grid of 10,000 sampled states, equally spaced, containing $\myset{C}$ defined with endpoints $\{ $ $[0.284, 0.284]^T,$ $  [0.284, 0.976]^T, $ $ [0.976, 0.284]^T, $ $ [0.976, 0.976]^T\}$. For the two scenarios presented in Section \ref{ssec:two tank example}, the sampling approach ensured that \eqref{eq:dcbf predictive} held over the grid search, which suggested that $h$ was an \emph{N-step} \RBF.

\bibliographystyle{IEEEtran}
\bibliography{bibliography}
\end{document}